%% file: ANOVA-GP.tex
\documentclass[review]{elsarticle}
\usepackage{moreverb}
\usepackage{hyperref}
\usepackage{graphicx,wrapfig}
\usepackage{framed,multirow}
\usepackage{float}
\usepackage{epstopdf}
\usepackage{array}
\usepackage{algorithm}
\usepackage{algpseudocode}
\usepackage{color}
\usepackage{amsmath}
\usepackage{txfonts}
\usepackage{afterpage}
\usepackage[margin=1in]{geometry}
\usepackage{mathrsfs}
\usepackage{tikz}
\usepackage{pstricks}
\usepackage{diagbox}
\usepackage{amssymb}
\usepackage{latexsym}
\usepackage{bm}
\usepackage{multicol}
\usepackage{subfigure}
\usepackage{bbm}

\font\cmmib=cmmib10 \font\msbm=msbm10

\newcommand{\bx}{\bm{x}}

\newcommand{\xij}{\xi^{(j)}}
\newcommand{\lt}{\left}
\newcommand{\rt}{\right}
\def\xict{\xi^{\,c,t}}
\def\pP{\mathcal{J}}
\def\pPp{\widehat{\mathcal{J}}}
\def\pL{\mathcal{M}}
\def\pC{C}
\def\pE{\mathbf{E}}
\def\pEp{\hat{\mathbf{E}}}
\font\cmmib=cmmib10 \font\msbm=msbm10 

\def\dsQ#1{\hbox{\cmmib Q}{}_{#1}}

\def\dsR{\hbox{{\msbm \char "52}}}
\def\ur#1{\tilde{u}_t(\xi_t^{#1})}
\def\urc#1#2{\tilde{u}_{t,\,#2}(\xi_t^{#1})}
\def\ure#1{\tilde{u}_{t}\left(\xi_t^{#1}\right)}
\def\urec#1#2{\tilde{u}_{t,\,#2}\left(\xi_t^{#1}\right)}
\def\mr#1{\overline{U}_t(\xi_t^{#1})}

\def\ume#1{\overline{u}_t\left(\xi_t^{#1}\right)}

\def\cov{\mathfrak{c}}
\def\mean{m'}
\def\var{v'}
\def\div{\nabla \cdot}

\renewcommand{\mathbf}{\boldsymbol}

\makeatletter
\def\ps@pprintTitle{%
	\let\@oddhead\@empty
	\let\@evenhead\@empty
	\let\@oddfoot\@empty
	\let\@evenfoot\@oddfoot
}
\makeatother

\begin{document}

\begin{frontmatter}

\title{ANOVA  Gaussian process modeling  for high-dimensional stochastic computational models}

\author[mymainaddress,mysecondaryaddress]{Chen Chen}
\ead{cchen24@umd.edu}

\author[mymainaddress]{Qifeng Liao\corref{mycorrespondingauthor}}
\cortext[mycorrespondingauthor]{Corresponding author}
\ead{liaoqf@shanghaitech.edu.cn}

\address[mymainaddress]{School of Information Science and Technology, ShanghaiTech University, Shanghai, China}
\address[mysecondaryaddress]{Department of Computer Science, University of Maryland, College Park, MD 20742,
	United States}

\begin{abstract}
In this paper we present a novel analysis of variance Gaussian process (ANOVA-GP) emulator for 
models governed by partial differential equations (PDEs) with high-dimensional random inputs. 
Gaussian process (GP) is a widely used surrogate modeling strategy, 
but it can become invalid when the inputs are high-dimensional. 
In this new ANOVA-GP strategy, high-dimensional inputs 
are decomposed into unions of local low-dimensional inputs,
and principal component analysis (PCA) is applied 
to provide dimension reduction for each ANOVA term. We then 
systematically build local GP models for PCA coefficients based on ANOVA decomposition 
to provide an emulator for the overall high-dimensional problem.
We present a general mathematical framework of ANOVA-GP, 
validate its accuracy and demonstrate its efficiency with numerical experiments.
\end{abstract}


\end{frontmatter}

\section{Introduction} \label{sec_Introduction}
\input{./Introduction}

\section{Problem setting}\label{sec_Problem_definition}
\input{./Problem_definition}

\section{ANOVA decomposition}\label{sec_ANOVA}
\input{./ANOVA}

\section{ANOVA Gaussian process modeling}
\label{sec_GP}
\input{./Manifold_learning_method}

\section{Numerical study}\label{sec_numer}
\input{./Numerical_study}

\section{Concluding remarks}\label{section_conclude}
Conducting dimension reduction is one of the fundamental concepts to develop efficient GP emulators 
for complex computational models with high-dimensional inputs and outputs. 
With a focus on adaptive ANOVA decomposition, this paper proposes a
novel ANOVA-GP strategy. 
In ANOVA-GP, the high-dimensional inputs 
are decomposed into a combination of low-dimensional local inputs through 
adaptive ANOVA decomposition, and PCA is applied on each ANOVA term \eqref{sum_adaptive_ANOVA}
to result in a reduced dimensional representation of the outputs.
Local GP models are built through active training with initial data obtained 
in the ANOVA decomposition procedure. 
Since each local input is low-dimensional and 
the resulting term in the ANOVA expansion has a small rank, GP emulation for 
each ANOVA term becomes less challenging compared with that for the overall problem \eqref{spde_l}--\eqref{spde_b}.
From numerical studies, it can be seen that a very small number of data points are required to build local GP models for each 
ANOVA term. It is also clear that for a given number of training data points, 
prediction errors of ANOVA-GP are smaller than the errors of the standard GP method.
In addition, the cost of ANOVA-GP for conducting predictions  
is cheaper than that of standard GP. 
Assuming that there are $N_{train}$ training data points for each ANOVA term in ANOVA-GP (see Algorithm \ref{AGP})
and there are $|\pP|$ ANOVA terms, the total number of training data points is $|\pP|N_{train}$.  
The cost of using GP models to make a single prediction is dominated by the cost of computing the inverse of
the covariance matrix (see \eqref{gp})---the main cost of ANOVA-GP is
then $O(|\pP|N^3_{train})$, and the main cost of the standard GP method
with $|\pP|N_{train}$ training data points is $O((|\pP|N_{train})^3)$
(which is larger than that of ANOVA-GP).
As in our ANOVA-GP setting, 
PCA is applied to conduct dimension reduction for the output space, 
and the number of training data points 
for different local GP models are all set to the same number $N_{train}$ 
(see Algorithm \ref{AGP}), which may not be optimal when the underlying 
problem has highly nonlinear structures. A possible solution is to apply nonlinear model reduction methods
and adaptive training procedures to result in different number of training data points for each 
ANOVA term, which will be the focus of our future work.

\bigskip
\textbf{Acknowledgments:}
This work is supported by the National Natural Science Foundation of China (No. 11601329).

\bibliography{chen,liao,tang}

\end{document}

%% file: Introduction.tex
During the last few decades there has been a rapid development in 
surrogate modeling for computational models governed by 
stochastic partial differential equations (PDEs). 
This explosion in interest has been driven by 
practical applications including  uncertainty quantification, 
shape and topological optimizations, and Bayesian inversions. 
In these applications, repeated simulations for parameterized PDE systems are demanded.
High-fidelity numerical schemes, which are also referred to as the simulators, 
can give accurate predictions for the outputs of these PDE systems, e.g.,
the finite element methods with a posteriori error bounds \cite{ainoden,Elman2014}.
However, the simulators are typically computationally expensive, especially when modeling complex science and engineering problems. 
In order to reduce the costs in these many-query problems of computational models, 
cheap surrogate models which are also called emulators, are actively developed  to replace 
the simulators. These include Gaussian process (GP) emulators 
\cite{kennedy2001bayesian,oakley-ohagan02,kennedy2006case,tagade2013cfd}, 
polynomial chaos surrogates \cite{ghanem2003sfem,xiu2002wiener} and reduced basis methods \cite{boybri10,elmanliao,cheroz14,chenjiang16}. 


The original GP emulator is to model the system output by a Gaussian process indexed by input parameters 
\cite{oakley-ohagan02}, 
which limits its application to high-dimensional problems.
In general, the computational models   
governed by stochastic PDEs have high-dimensional inputs and outputs. 
There are always a large number of input parameters, 
when modeling complex problems, for example, 
models with inputs described by rough random processes with short correlation lengths.
The standard outputs of the PDE systems are the spatial fields, 
and when a fine resolution representation is required, the outputs need to be high-dimensional to capture 
detailed local information. 
This kind of high-dimensional problems currently gains a lot of interests,
and new GP emulators are actively developed.
These new GP methods usually focus on either a high-dimensional input space 
or a high-dimensional output space, 
and propose  dimension reduction techniques for the corresponding high-dimensional space. 
In \cite{higdon2008computer}, principal component analysis (PCA) is applied to the output space to result in an efficient GP emulator 
for models with high-dimensional outputs.  
In \cite{ma2011kernel,xing2016manifold}, novel kernel principal component analysis is developed to perform dimension reduction for the output space.  
In addition, an active data selection method is developed to build GP surrogates for PCA coefficients 
in \cite{guo2018reduced}. 
For problems with high-dimensional inputs, GP with built in active subspace dimension reduction is proposed in \cite{tripathy2016gaussian}. 


We focus on the challenging situation that both inputs and outputs are high-dimensional.
A main challenge here is that difficulties caused by high-dimensional inputs and outputs
are typically coupled. As  discussed in \cite{elmanliao}, high-dimensional inputs can lead to large ranks in the 
output space, and direct PCA for the output space can consequently become inefficient. 
To decouple the difficulties, 
we propose a novel analysis of variance (ANOVA) based Gaussian process method (ANOVA-GP). 
In this ANOVA-GP emulator, the high-dimensional parameter space is decomposed into a union of
low-dimensional spaces through an adaptive ANOVA procedure. 
PCA is conducted locally on ANOVA terms associated with these low-dimensional parameter spaces. 
After that,  local GP models are built for PCA coefficients. 
Since the local inputs are low-dimensional, efficient PCA can be achieved
and a small number of training data points are required to result in 
accurate local GP models. In addition, 
we note that a Bayesian smoothing spline ANOVA Gaussian process framework is developed for model calibration 
with categorical parameters \cite{storlie2015calibration}, but the novelty of our ANOVA-GP lies on adaptive construction procedures 
for hierarchical GP models for high-dimensional (noncategorical) parameters.

An outline of the rest of the paper is as follows.
Section \ref{sec_Problem_definition} sets the problem, 
and section \ref{sec_ANOVA} gives a detailed discussion of the ANOVA decomposition.
In section \ref{sec_GP}, we first discuss PCA for each ANOVA term and active training for each local GP model,
and next present our novel overall ANOVA-GP emulator.
Numerical results are discussed in Section \ref{sec_numer}. 
Second \ref{section_conclude} concludes the paper.

%% file: Problem_definition.tex
Let $D$ denote a physical domain (in $\dsR^2$ or $\dsR^3$) which is bounded, 
connected and with a polygonal boundary $\partial D$. 
Suppose $\xi=[\xi_1,\dots,\xi_m]^T$ is a $m$-dimensional vector which collects a finite number of independent random variables 
and the probability density function of $\xi$ is denoted by $\pi(\xi)$. 
Without loss of generality, we further assume that $\xi$ has a bounded and connected support $I^m$, where $I$ is a real closed interval. In this paper, we consider physical problems governed by  PDEs over the physical domain $D$ and boundary conditions 
on the boundary $\partial D$, which can be stated as: find a stochastic function  
$u_{sol}(x, \xi): D \times I^m \rightarrow \mathbb{R}$, such that 
\begin{eqnarray}
\mathcal{L}\left(x, \xi;\ u_{sol}\left(x, \xi \right) \right) = f\left(x, \xi\right) & \forall \left(x,\xi \right) \in D\times I^m,\label{spde_l}\\
\mathcal{B}\left(x, \xi;\ u_{sol}\left(x, \xi\right)\right) = g\left(x, \xi\right)& \forall \left(x,\xi \right) \in \partial D\times I^m,\label{spde_b}
\end{eqnarray}
where  $\mathcal{L}$ is a partial differential operator and $\mathcal{B}$ is a boundary operator,
both of which can have random coefficients.
Given a realization of $\xi$ which is denoted by $\xij$ ($j\in \mathbb{N}$), 
a simulator (e.g., the finite element method~\cite{Elman2014}) can provide approximate values of 
$u_{sol}(x, \xi)$ on given physical grid points,
which result in a high-dimensional output. We denote this output as 
\begin{equation}
y^{(j)}: = u \left(\xij\right):= \left[u_{sol}\left(x^{\left(1\right)}, \xij\right),\dots,u_{sol}\left(x^{\left(d\right)}, \xij\right)\right]^T \in \mathbb{R}^d,\label{discrete representation}
\end{equation} 
where $d$ is the number of grid points (or the finite element degrees of freedom) and $x^{\left(k\right)},k=1,\dots,d$ 
are the locations of the grid points. 
Letting $\mathcal{O} \subset \mathbb{R}^d$ denote the manifold consisting of $u\left(\xi\right)$ 
associated with all realizations of $\xi$,
a simulator can be viewed as a mapping 
$\chi: I^m \rightarrow \mathcal{O}$.
The inputs and the outputs of $\chi$ are both high-dimensional in this general setting, 
which causes difficulties for applying traditional GP methods.
For this purpose, we in this work provide a novel ANOVA-GP surrogate for $\chi$,
where ANOVA decomposition is conducted to decompose the high-dimensional 
inputs into a union of low-dimensional local inputs. For each local input, PCA is applied to
result in a reduced dimensional representation of the corresponding local output.
After that, local GP models are built for the PCA coefficients. 
The next section is to review the ANOVA decomposition following the presentation in 
\cite{rabitz1999efficient, rabitz1999general, gao2010anova,ma2010adaptive,zhachokar12,yang2012adaptive,hesthaven2016use},
while PCA for the outputs and our overall ANOVA-GP strategy are presented in section \ref{sec_GP}.

%% file: ANOVA.tex
Let $\mathcal{P}$ be the set consisting of  coordinate indices $\{1,2,\dots, m\}$. 
Any non-empty subset $t\subseteq \mathcal{P}$ is referred to as an ANOVA index, and
the elements of $t$ are sorted in ascending order, while its cardinality is denoted by $\vert t\vert$. 
For a given $t$,
let $\xi_t$ denote a $|t|$-dimensional vector that includes components of the vector $\xi \in I^{m}$ indexed by $t$. 
For example, if $t = \{1,3,4\}$, then $|t| = 3$ and $\xi_t = [\xi_1, \xi_3, \xi_4]^{T}\in I^3$. Letting $\mathrm{d}\mu$ denote a given probability measure on $I^m$, 
the ANOVA decomposition of the simulator output  $u(\xi)$ of the problem~\eqref{spde_l}--\eqref{spde_b} can be expressed as
\begin{equation}
u\left(\xi\right) = \sum\limits_{t \subseteq \mathcal{P}} u_t\left(\xi_t\right). \label{comfunc}
\end{equation}
In \eqref{comfunc}, each term on the right hand side is defined recursively through
\begin{equation}
u_t\left(\xi_t\right) = \int_{I^{m-|t|}}u\left(\xi\right)\mathrm{d}\mu\left(\xi_{\mathcal{P}\backslash t}\right) - \sum\limits_{w \subset t}u_w\left(\xi_w\right),\label{comfunc1}
\end{equation}
starting with
\begin{equation}
u_\emptyset = \int_{I^m}u\left(\xi\right)\mathrm{d}\mu\left(\xi\right), \label{comfunc2}
\end{equation}
where $\mathrm{d}\mu\left(\xi_{\mathcal{P}\backslash t}\right):=\prod_{i\in \mathcal{P}\backslash t}\mathrm{d}\mu\left(\xi_i\right)$,
since $\{\xi_i\}^m_{i=1}$ are assumed to be independent. 
Note that $u(\xi)$ is a vector and integrals involving them (e.g.,  \eqref{comfunc1} and \eqref{comfunc2}) are defined componentwise.   
In this paper, we call $u_t\left(\xi_t\right)$ a $|t|$-th order ANOVA term and $t$ a $|t|$-th order index. 

When the ordinary Lebesgue measure is used in~(\ref{comfunc1})--(\ref{comfunc2}), 
\eqref{comfunc} is referred to as the classic ANOVA decomposition,
and each expansion term is  
\begin{equation}
u_t\left(\xi_t\right) = \int_{I^{m-|t|}}u\left(\xi\right)\prod_{i\in \mathcal{P}\backslash t}\mathrm{d}\xi_i - \sum\limits_{w \subset t}u_w\left(\xi_w\right),\label{comfunc3}
\end{equation}
and\begin{equation}
u_\emptyset = \int_{I^m}u\left(\xi\right)\mathrm{d}\xi.\label{comfunc4}
\end{equation}
Computing each term \eqref{comfunc3}  in the classic ANOVA decomposition requires 
computing integrals over $I^{m-|t|}$. When $|t|$ is small, $I^{m-|t|}$ has a high dimensionality,
and computing integrals over it is expensive. To alleviate this difficulty, anchored ANOVA
methods \cite{sobol2003theorems} are developed, and are reviewed as follows.

\subsection{Anchored ANOVA decomposition}
As discussed in \cite{sobol2003theorems,ma2010adaptive,yang2012adaptive}, the idea of anchored 
ANOVA decomposition is to replace the Lebesgue measure used in 
 \eqref{comfunc3}--\eqref{comfunc4} by the Dirac measure 
\begin{equation}
\mathrm{d}\mu\left(\xi\right) := \delta\left(\xi - c\right)\mathrm{d}\xi = \prod_{i=1}^m\delta\left(\xi_i-c_i\right)\mathrm{d}\xi_i,
\end{equation}
where $c = [c_1, c_2, \dots, c_m]^T\in I^m$ is a given anchor point. With the Dirac measure, 
each term in \eqref{comfunc1} is
\begin{equation}
u_t\left(\xi_t\right) = u\left(\xict\right) - \sum\limits_{w \subset t}u_w\left(\xi_w\right),\label{comfunc5}
\end{equation}
where the initial term is set to $u_\emptyset = u\left(c\right)$ and 
$\xict:=[\xict_1,\ldots,\xict_m]^T\in I^m$ is defined through
\begin{eqnarray}
\xict_i&:=&\left\{\begin{array}{ll} 
                     c_i & \textrm{for } i\in \{1,\ldots,m\}\setminus t\\
                     \xi_i & \textrm{for } i \in t 
                     \end{array}\right. .\label{xict}
\end{eqnarray}
The anchored ANOVA decomposition expresses the simulator output $u(\xi)$ 
by the knowledge of its values on lines, planes and 
hyper-planes passing through the anchor point $c$ \cite{rabitz1999efficient}.  
Here comes a natural question that how to choose the anchor point. 
Generally, the anchor point can be chosen arbitrarily since the ANOVA decomposition~(\ref{comfunc}) is always exact. 
However, an appropriately chosen anchor point enables the decomposition to give an accurate approximation with
a small number of expansion terms \cite{sobol2003theorems, wang2008approximation}, 
which give computational efficiency (the selection procedure of ANOVA terms is discussed in the next section).    
In ~\cite{sobol2003theorems, wang2008approximation},
it is shown that a good choice is the input sample point where the corresponding output sample equals or is close to 
the mean of the output. However, the mean of the output is not given a priori in our setting,
and it is not trivial to find the input sample point which gives an output sample close to the mean of output.
As shown in \cite{xu2004generalized,gao2010anova}, an optimal choice is mean of the input,
and we use this choice of the anchor point for all numerical studies in this paper.

It is clear that  the whole index set $\{t\ |\ t \subseteq \mathcal{P}\}$ contains a large number of terms when $m$ is large,
and  especially, the $m$-th order index $t=\{1,\ldots,m\}$ is included,
which causes challenges to compute  the right hand side of \eqref{comfunc}. 
However, in practical computation, not all expansion terms in \eqref{comfunc} need to be computed---only low order ANOVA terms 
are typically considered to be active and need to be computed.
Denoting a selected index set by  
$\pP$ which is a subset of the whole index set $\{t\ |\ t \subseteq \mathcal{P}\}$,
an approximation of the solution $u\left(\xi\right)$ is written as
\begin{equation}
u\left(\xi\right) \approx u_{\pP}\left(\xi\right) := \sum\limits_{t \in \pP}u_t\left(\xi_t\right),\label{sum_adaptive_ANOVA}
\end{equation}
where $u_t\left(\xi_t\right)$ is defined in (\ref{comfunc1}).
Next, we review the adaptive construction procedure for the index set $\pP$ following \cite{ma2010adaptive}.

\subsection{Adaptive index construction} \label{section_anova_adaptive}
For each $i=0,\ldots,m$, the set consisting of selected $i$-th order indices is denoted by $\mathcal{J}_i$,
while $\mathcal{J}=\cup^m_{i=0} \mathcal{J}_i$.
For the zeroth order index, we set $\mathcal{J}_0 = \{\emptyset\}$ and $|\emptyset|=0$, 
and  $u(c)$ is computed using a given simulator (e.g., the finite element method).
Supposing that $\pP_i$ is known for a given order $0 \leq i \leq m-1$,  $\mathcal{J}_{i+1}$ 
is constructed based on $\pP_i$ as follows. 
First, a candidate index set $\pPp_{i+1}$ is constructed as 
\begin{align}
\pPp_{i+1} &:= \left\{t\ \big|\ \vert t\vert = i + 1, {\rm and\  any\ }s \subset t\ {\rm with\ } |s|=i\ {\rm satisfies\ }s \in \mathcal{J}_i\right\}.\label{adaptive}
\end{align}
For each $t \in \widehat{\mathcal{J}}_{i+1}$, the contribution weight of $u_t\left(\xi_t\right)$ is 
defined as 
\begin{equation}
\gamma_t := \frac{\left\|\pE\left(u_t\left(\xi_t\right)\right)\right\|_{L_2}} {\left\|\sum_{s \in \pP_0\cup\cdots\cup \pP_{|t|-1}} \pE\left(u_s\left(\xi_s\right)\right)\right\|_{L_2}}, \label{weights}
\end{equation}
which measures the relative importance of the index $t$ \cite{ma2010adaptive}.  
In \eqref{weights}, $\| u_t(\xi_t) \|_{L_2}$ is the  functional $L_2$ norm of the approximation 
function associated with $u_t(\xi_t)$ (e.g., the finite element approximation function with coefficients defined by $u_t(\xi_t)$
\cite{Elman2014}),  
and $\pE\left(u_t\left(\xi_t\right)\right)$ denotes the mean function of $u_t$ that is defined as
\begin{equation}
\pE\left(u_t\left(\xi_t\right)\right) = \int_{I^{|t|}}u_t\left(\xi_t\right)\pi_t\left(\xi_t\right)\mathrm{d}\xi_t,
\end{equation}
where $\pi_t(\xi_t)$ is the marginal probability density function of $\xi_t$. This mean function can be approximated using the Clenshaw-Curtis tensor quadrature rule~\cite{novak1996high, trefethen2008gauss,yang2012adaptive}, i.e.,
\begin{equation}
\pEp\left(u_t\left(\xi_t\right)\right) := \sum_{\xi_t^{\left(k\right)} \in \Xi_t} u_t\left(\xi_t^{(k)}\right)\pi_t\left(\xi_t^{(k)}\right)w\left(\xi_t^{(k)}\right),\ k=1,2,\dots,\vert \Xi_t\vert, \label{sgp}
\end{equation}
where $\Xi_t$ contains the Clenshaw-Curtis tensor quadrature points, $\{w(\xi_t^{(k)})\}$ for $k=1,\ldots,|\Xi_t|$ 
are the corresponding weights, and $|\Xi_t|$ is the size of $\Xi_t$. 
After that, the set $\mathcal{J}_{i+1}$ is formed through the $(i+1)$-th order indices with $\gamma_t \geq tol_{index}$,
i.e., $\pP_i:=\{t\,| \,t\in\widehat{\pP}_i \textrm{ and }\gamma_t \geq tol_{index}\}$, 
where $tol_{index}$ is a given tolerance.  
This hierarchical construction  procedure stops when $\widehat{\mathcal{J}}_{i+1} = \emptyset$. 

The above procedure to adaptively select ANOVA terms is summarized  in Algorithm~\ref{AA}. 
\begin{algorithm}[htpb!]
	\caption{The adaptive anchored ANOVA decomposition.}
	\label{AA}
	\begin{algorithmic}[1] 
		\Require   A simulator for \eqref{spde_l}--\eqref{spde_b} and the probability density function of $\xi$.
		\State Initialize: $\pP = \{\emptyset\}$, $\widehat{\mathcal{J}}_1 = \{1,2\dots,M\}$ and $i=1$.
		\State Compute $u(c)$ through the given simulator, where $c$ is the anchor point. 
		\While {$\widehat{\mathcal{J}}_{i} \neq \emptyset$}
		\State Set $\pP_i = \emptyset$. 
		\For {$t\in \pPp_{i}$}
		\State Setup the Clenshaw-Curtis tensor quadrature points $\Xi_t=\left\{\xi_t^{(1)},\ldots,\xi_t^{(|\Xi_t|)}\right\}$ 
		and weights $\left\{w\left(\xi_t^{(k)}\right)\right\}$ for $k=1,\ldots,|\Xi_t|$. 
		\State Compute $ \gamma_t:= \frac{\left\|\pEp\left(u_t\left(\xi_t\right)\right)\right\|_{L_2}}
		{\left\|\sum_{s \in\pP}\pEp\left(u_s\left(\xi_s\right)\right)\right\|_{L_2}}$, 
		where $\pEp$ is defined in \eqref{sgp} and  $u_t(\xi_t)$
		is computed using \eqref{comfunc5} with the simulator. 
		\If {$\gamma_t > tol_{index}$}
		\State Update $\mathcal{J}_i = \mathcal{J}_i \cup \{t\}$. 
		\EndIf
		\EndFor
		\State Update $\mathcal{J} = \mathcal{J} \cup \mathcal{J}_i$. 
		\State Construct $\widehat{\mathcal{J}}_{i+1} = \left\{t\ \big|  \ |t| = i + 1, {\rm and\ any\ }s \subset t\ {\rm with\ } |s|=i\ {\rm satisfies\ }s \in \mathcal{J}_i\right\}$. 
		\State Update the ANOVA order: $i=i+1$.
		\EndWhile
		\Ensure An effective index set $\pP$ and 
                               data sets $\Theta_t:=\left\{y_t^{(k)}=u_t\left.\left(\xi_t^{(k)}\right)\, \right|\, \xi_t^{(k)}\in \Xi_t, \textrm{ and } k=1,\ldots,|\Xi_t| \right\}$ for  $t\in\pP$.
	\end{algorithmic}
\end{algorithm}

%% file: Manifold_learning_method.tex
In this section, our novel ANOVA Gaussian process (ANOVA-GP) modeling strategy is presented.
This new strategy is based on building GP models for each ANOVA term. 
It is clear that, the dimension of each ANOVA term in \eqref{sum_adaptive_ANOVA} 
is the same as that of the simulator output, e.g., the finite element 
degrees of freedom, which is high-dimensional. As discussed in section~\ref{sec_Introduction}, it is challenging to apply 
 standard GP models for problems with high-dimensional outputs. To result in a reduced dimensional representation
 of the output, we apply the principal component analysis (PCA) \cite{jolliffe2011principal,vms2016gpca} for each ANOVA term. 
 After that, based on the data sets obtained in the ANOVA decomposition step (see section~\ref{section_anova_adaptive}), 
 an active training procedure is developed to construct the GP models for each PCA mode. 
 Our overall ANOVA-GP procedure is summerized at the end of this section.

\subsection{Principal component analysis}\label{4.1}
The principal component analysis \cite{jolliffe2011principal} is to find
subspaces in which observed data can be approximated well. The basis vectors of these subspaces are called
the principle components, which are also referred to as proper orthogonal decomposition bases 
\cite{holmes96,benner15}. 
In this work, PCA is applied to obtain reduced dimensional representations for each ANOVA term $u_t(\xi_t)$ 
in \eqref{sum_adaptive_ANOVA}.
To conduct PCA for  $u_t(\xi_t)$, a data set consisting of samples of $u_t(\xi_t)$ is required.
In this section, the data set of $u_t(\xi_t)$ is generically denoted by 
$\vartheta_t:=\{y^{(j)}_t \, | \, y^{(j)}_t=u_t(\xi_t^{(j)})  \in \dsR^{d} \textrm{ and } j=1,\ldots,N\}$, where $N$ 
denotes the size of $|\vartheta_t|$. 
Note that the ANOVA decomposition procedure (Algorithm \ref{AA}) gives  
a data set  $\Theta_t:=\{y_t^{(k)}=u_t(\xi_t^{(k)})\, |\, \xi_t^{(k)}\in \Xi_t, \textrm{ and } k=1,\ldots,|\Xi_t| \}$
for each $t\in \pP$. This data set $\Theta_t$ can be used as an initial choice for $\vartheta_t$ to conduct PCA,  
while an active training procedure based on our new selection criterion  provides additional sample points, 
 which is discussed in 
section \ref{sec_at}. 

For each data set $\vartheta_t$ for $t\in \pP$,
the first step of PCA is to normalize the sample mean as follows
\begin{itemize}
	\item[1)] $\displaystyle \mu_t=\frac{1}{N}\sum_{j=1}^{N}y_t^{(j)}$, for $j=1,\dots,N$,
	\item[2)] $\displaystyle y_t^{(j)}\leftarrow y_t^{(j)}-\mu_t$.
\end{itemize} 
After that, the empirical covariance matrix is assembled
\begin{equation*}
\Sigma = \frac{1}{N}\sum_{j=1}^{N}y_t^{(j)}\left(y_t^{(j)}\right)^T.
\end{equation*}
The eigenvalues and the eigenvectors of $\Sigma$ are denoted by $\lambda_1\geq \ldots \geq \lambda_N $ 
and $v_1,\ldots, v_N$ respectively.
For a given tolerance 
$tol_{pca}$, 
the first 
$R$ 
eigenvectors 
$\{v_1,\ldots,v_R \}$ satisfying 
${\sum_{j=1}^{R}\lambda_j}/{\sum_{j=1}^{N}\lambda_j}>1-tol_{pca}$
but ${\sum_{j=1}^{R-1}\lambda_j}/{\sum_{j=1}^{N}\lambda_j}\leq 1-tol_{pca}$,
are referred to as the principle components.
In addition, $V_t:=[v_1,\ldots,v_R]$ denotes the matrix collecting the principle components. 
Details of PCA for each ANOVA term $u_t\left(\xi\right)$,  $t\in \pP$ are summarized in Algorithm \ref{PCA}.

With the principle components, each ANOVA term $u_t(\xi_t)\in \dsR^d$
for an arbitrary realization of $\xi_t$  can be approximated as:
\begin{equation*}
u_t(\xi_t) \approx  V_t \ure{}+\mu_t,
\end{equation*}
where 
\begin{equation}
\label{pca_ap}
\ure{}=[\urec{}{1},\ldots,\urc{}{R}]^T \in \dsR^R \textrm{, }\quad \urec{}{r}:=v_r^T(u_t(\xi_t)-\mu_t) \textrm{ for } r=1,\ldots,R.
\end{equation}
In the following, $\ur{}$ is referred to as  
the principle component representation (PC representation) of $u_t(\xi_t)$.

\begin{algorithm}[]
	\caption{Principal component analysis for each ANOVA term $u_t\left(\xi_t\right)$,  $t\in \pP$.}
	\label{PCA}
	\begin{algorithmic}[1]
		\Require  A data set $\vartheta_t:=\left.\left\{y^{(j)}_t \, \right| \, y^{(j)}_t=u_t\left(\xi_t^{(j)}\right)  \in \dsR^{d} \textrm{ and } j=1,\ldots,N\right\}$.
		\State Compute the sample mean:  $\mu_t=\frac{1}{N}\sum_{j=1}^{N}y_t^{(j)}$.
		\State Normalize the data: $y_t^{(i)}\leftarrow y_t^{(i)}-\mu_t$ for $j=1,\ldots,N$.
		\State Construct the covariance matrix: $\Sigma = \frac{1}{N}\sum_{j=1}^Ny_t^{(j)}\left(y_t^{(j)}\right)^T$.
		\State Compute eigenpairs $(\lambda_k, v_k)$ of $\Sigma$, where $k=1,\ldots,N$ and $\lambda_1\geq \ldots \geq \lambda_N$.
		\State Select the first $R$ eigenvectors $\{v_1,\ldots,v_R\}$ such that $ \frac{\sum_{j=1}^{R}\lambda_j}{\sum_{j=1}^{N}\lambda_j}	  > 1-tol_{pca}$, but  $\frac{\sum_{j=1}^{R-1}\lambda_j}{\sum_{j=1}^{N}\lambda_j}\leq 1-tol_{pca}$. 
		\State Compute  
			$\ure{(j)}=\left[\urec{(j)}{1},\ldots,\urec{(j)}{R}\right]^T \leftarrow \lt[v_1^Ty_t^{(j)},\dots,v_R^Ty_t^{(j)}\rt]^T$, where $j=1,\dots,N$.   
		\State Construct data vectors for PCA coefficients: $\alpha_{t,r} = \left[\urec{(1)}{r}, \dots, \urec{(N)}{r}\right]^T$ 
		for $r=1,\ldots,R$.
		\Ensure the sample mean $\mu_t$,  
		the principle component matrix $V_t:=[v_1,\ldots,v_R]$,  
                 eigenvalues $\lambda_r$	and data vectors  $\alpha_{t,r}$ for $r=1,\ldots,R$. 
	\end{algorithmic}
\end{algorithm} 

\subsection{Gaussian process regression with active training}\label{sec_at}
In this section for each $t\in \pP$, following the active data selection method developed in \cite{guo2018reduced}, 
a Gaussian process modeling strategy with 
active training is proposed for 
each PC representation $\ur{}=[\urc{}{1},\ldots,\urc{}{R}]^T\in \dsR^R$ (see \eqref{pca_ap}). 
Due to the compression obtained through PCA, the dimension $R$ is typically very small
and independent of the dimension of the simulator output (e.g., the finite element degrees of freedom). 
So, it is computationally feasible to construct GP models for each ANOVA term independently.

A Gaussian process is a collection of random variables, 
and any finite combinations of these random variables 
are joint Gaussian distributions. 
In our setting, 
for each realization of $\xi_t$,  $\urc{}{r}$ is considered to be a random variable in 
a Gaussian process. 
Following the presentation in \cite{xing2016manifold},
each of the prior GP models is denoted  
by $\urc{}{r} \sim \mathcal{GP}(m(\xi_t), \cov(\xi_t, \xi'_t))$
 where $m(\cdot)$ is the mean function and $\cov(\cdot, \cdot)$ is the covariance function 
 of the Gaussian process $\mathcal{GP}$ that needs to be trained.  
The Gaussian process is specified by its mean function and covariance 
function~\cite{rasmussen2004gaussian}. 
In this work, the mean function is set to $m(\xi_t) = 0$, 
and the covariance function is set to a noisy squared exponential function
\begin{equation}
\cov\lt(\xi_t, \xi'_t\rt) = \rho_1^2 \cdot \mathrm{exp}\lt(-\lt(\xi_t-\xi'_t\rt)^T\mathrm{diag}\lt(\ell_1,\dots,\ell_M\rt)^{-1}\lt(\xi_t-\xi'_t\rt) / 2\rt) + \rho_2^2\cdot\delta\lt(\xi_t, \xi'_t\rt). \label{cov}
\end{equation}
The last term in~(\ref{cov}) is called `jitter'~\cite{andrianakis2012effect},  $\delta(\xi_t, \xi'_t)$ is a Kronecker delta function 
which is one if $\xi_t = \xi'_t$ and zero otherwise,
and $\mathrm{diag}(\ell_1,\dots,\ell_M)$ is a diagonal matrix. The hyperparameters $\ell_1,\dots,\ell_M$ and $\rho_1^2, \rho_2^2$ are square correlation lengths and signal variances respectively. 
Denoting $\beta = [\ell_1,\dots,\ell_M, \rho_1, \rho_2]^T$, the hyperparameters 
can be determined through minimizing the following negative log marginal likelihood $\pL(\beta)$:
\begin{equation}
\pL\lt(\beta\rt) = -\log p\lt(\alpha_{t,r} |\, \beta\rt) = \frac{1}{2} \log \mathrm{det}\ \left(\pC\lt(\beta\rt)\right) + \frac{1}{2}\alpha_{t,r}^T\pC^{-1}\lt(\beta\rt)\alpha_{t,r} + \frac{N}{2}\log\lt(2\pi\rt), \label{loglike}
\end{equation}
where $\alpha_{t,r} = [\urc{(1)}{r}, \dots, \urc{(N)}{r}]^T$ is the training target and $\pC(\beta)$ is the covariance matrix with entries 
$\pC(\beta)_{jk} = \cov(\xi_t^{(j)},\xi_t^{(k)})$ for $j,k=1,\dots,N$. 
Minimizing $\pL(\beta)$ is a non-convex optimization problem~\cite{snelson2007flexible}, 
and we use the MATLAB toolbox \cite{rasmussen2010gaussian} to solve it,
where conjugate gradient methods are included~\cite{moller1993scaled}.

Once the hyperparameters are determined, from the joint distribution of  $\urc{}{r}$ and $\alpha_{t,r}$, 
the conditional predictive distribution for any arbitrary realization of $\xi_t$ is:
\begin{equation}
\urec{}{r} \ |  \alpha_{t,r}, \beta \sim U_{t,r}(\xi_t):=\mathcal{GP}\lt(\mean_r(\xi_t,\beta),\var_r\lt(\xi_t,\beta\rt)\rt), \label{gp}
\end{equation}
where $\mean_r(\xi_t,\beta) = c_*^T\pC(\beta)^{-1}\alpha_{t,r}$, 
$\var_r\lt(\xi_t,\beta\rt) = \cov\lt(\xi_t, \xi_t\rt) - c_*^T\pC\lt(\beta\rt)^{-1}c_{*}$,
and $c_* = [\cov(\xi_t, \xi_t^{(i)}),\dots,\cov(\xi_t, \xi_t^{(N)})]^T$ 
(see \cite{rasmussen2004gaussian}). 
Collecting the GP models for each PCA mode, the GP model for the overall PC representation for $\ure{}$ \eqref{pca_ap}
is denoted by $U_t(\xi_t):=[U_{t,1}(\xi_t),\ldots,U_{t,R}(\xi_t)]^T$ for each $t\in \pP$. 
For a given realization of $\xi_t$, the predictive mean of $U_t(\xi_t)$ is  
\begin{equation}
\mr{}:=\left[\mean_1\left(\xi_t,\beta\right),\dots,\mean_R\left(\xi_t,\beta\right)\right]^T. 
\end{equation}
With the principal components and the GP models for PC representations,
each ANOVA term $u_t(\xi_t)$ can be approximated as the following {\em local GP model} 
(the setting of a global GP model to approximate the overall problem \eqref{spde_l}--\eqref{spde_b} 
is discussed in section \ref{sec_numer}),
\begin{equation}
 \hat{u}_t(\xi_t):=V_tU_t(\xi_t)+\mu_t, \label{local-gp}
\end{equation}
where $V_t$ is the matrix consisting of the principal components and $\mu_t$
is the sample mean generated by Algorithm~\ref{PCA}.
The predictive mean of the local GP model model is 
\begin{equation}
 \ume{}:=V_t\mr{}+\mu_t. \label{local-gp-mean}
\end{equation}

It is clear that building a local GP model \eqref{local-gp} involves two main procedures: PCA and 
GP regression for each PCA mode. Both of these procedures are determined by the data set $\vartheta_t$ (the input of Algorithm \ref{PCA}). 
Our strategy is to use the data set $\Theta_t$ generated by the ANOVA decomposition step (Algorithm \ref{AA}) 
as an initial input data set to conduct PCA and to build the GP model for each PCA mode, i.e., initially set $\vartheta_t:=\Theta_t$. 
After that,  following \cite{guo2018reduced}, 
an active training method is developed to argument the training data set $\vartheta_t$ gradually 
to result in an accurate local GP model for $u_t(\xi_t)$, which proceeds as follows.
First, a candidate parameter sample set $\Psi$ is constructed using realizations of $\xi_t$ (different from the quadrature points 
$\Xi_t$ for 
ANOVA decomposition in Algorithm \ref{AA}). 
Second, for each sample in $\xi_t\in \Psi$, a variance indicator of the current GP model is computed as
\begin{equation}
\displaystyle \tau\lt(\xi_t\right):= \sum_{r=1}^{R}{\lambda_r\var_r\lt(\xi_t,\beta\rt)}\left/{\sum_{r=1}^{R}\lambda_r}\right., 
\end{equation}
where $\lambda_1,\ldots,\lambda_R$ are the eigenvalues generated in PCA with the current input data set, 
and $\var_r(\xi_t,\beta)$ is the variance of the current GP model for each 
PCA mode (see \eqref{gp}). Third, the input sample point with the largest variance indicator value   
$\xi_t^* = \mathrm{max}_{\xi_t\in \Psi} \tau(\xi_t)$ is selected to augment the input data set $\vartheta_t$,
and the local GP model is reconstructed with this augmented data set.
The second and the third steps are repeated until $\vartheta_t$ 
includes  $N_{train}$ data points, where $N_{train}>|\Xi_t|$ is a given number.
Details of this active training procedure are shown  in Algorithm~\ref{data_selection}. 

\begin{algorithm}[]
	\caption{Local GP modeling with active training for each ANOVA term $u_t\left(\xi\right)$,  $t\in \pP$.}
	\label{data_selection}
	\begin{algorithmic}[1]
		\Require  The number of training points $N_{train}$, and an initial training data set $\Theta_t$.
             \State Initialize a candidate set $\Psi$ consisting of realizations  of $\xi_t$, with size $|\Psi|>N_{train}$.  
             \State Initialize the training data set $\vartheta_t:=\Theta_t$. 
             \State Use Algorithm~\ref{PCA} with $\vartheta_t$ to obtain  
                  the sample mean $\mu_t$,  
	         the principle component matrix $V_t:=[v_1,\ldots,v_R]$, 
                 the eigenvalues $\lambda_r$ and the data vectors  $\alpha_{t,r}$ for $r=1,\ldots,R$. 
		\State Train GP models $U_{t,r}:=\mathcal{GP}\lt(\mean_r(\xi_t,\beta), \var_r\lt(\xi_t,\beta\rt)\rt)$ based on 
		the training data vectors $\alpha_{t,r}$ for each 
		PCA mode $r=1,\ldots,R$  (see \eqref{loglike}--\eqref{gp}). 
		\If {$|\vartheta_t| < N_{train}$}
		\State For each $\xi_t \in \Psi$, compute the variance 
		indicator $ \tau\lt(\xi_t\right):= \sum_{r=1}^{R}{\lambda_r\var_r\lt(\xi_t,\beta\rt)}\left/{\sum_{r=1}^{R}\lambda_r}\right.$.  
		\State Find $\xi_t^{*} = \mathrm{\arg\ max}_{\xi_t\in \Psi}\tau\lt(\xi_t\rt)$.
		\State Update the training data set: $\vartheta_t = \vartheta_t \cup \left\{u_t\lt(\xi_t^{*}\rt)\right\}$. 
		\State Remove $\xi_t^*$ from the candidate set: $\Psi = \Psi \setminus \xi_t^{*}$.
		\State Go to line 3. 
		\EndIf
		\State Construct the GP model for the PC representation: $U_{t}=[U_{t,1},\ldots,U_{t,R}]^T\in \dsR^R$ with 
		$U_{t,r}:=\mathcal{GP}\lt(\mean_r(\xi_t,\beta), \var_r\lt(\xi_t,\beta\rt)\rt)$ for $r=1,\ldots,R$. 
		\State Construct the local GP model:  $\hat{u}_t(\xi_t)=V_tU_t(\xi_t)+\mu_t$. 
		\Ensure 
		The local GP emulator $\hat{u}_t(\xi_t)$. 
	\end{algorithmic}
\end{algorithm}

\subsection{Overall ANOVA-GP model}\label{section_alg} 
With a given simulator for the problem \eqref{spde_l}--\eqref{spde_b}, 
our overall ANOVA-GP modeling proceeds as the following three main steps. 
First, ANOVA decomposition is conducted using Algorithm~\ref{AA}, which gives an effective index set $\pP$
and initial training data sets  
$\Theta_t:=\{y_t^{(k)}=u_t(\xi_t^{(k)})\, |\, \xi_t^{(k)}\in \Xi_t, \textrm{ and } k=1,\ldots,|\Xi_t| \}$ for $t\in\pP$.
After that, the local GP modes $\hat{u}_t(\xi_t)$ for each ANOVA index $t\in\pP$
are built using Algorithm~\ref{data_selection}. 
Finally, the overall ANOVA-GP model is assembled as
\begin{equation}
 \hat{u}_{\pP}\left(\xi\right):= \sum_{t \in \pP} \hat{u}_t\left(\xi_t\right). \label{anova-gp}
\end{equation}
For each realization  of $\xi$,
the predictive mean of the ANOVA-GP model is, 
\begin{equation}
 \overline{u}_{\pP}\left(\xi\right):= \sum_{t \in \mathcal{J}} \overline{u}_t\left(\xi_t\right), \label{anova-gp-mean}
\end{equation}
where $\overline{u}_t (\xi_t )$ is the predictive mean of the local GP model defined in \eqref{local-gp-mean}.
This ANOVA-GP modeling procedure is summarized in Algorithm \ref{AGP}.

\begin{algorithm}[]
	\caption{ANOVA-GP modeling.}
	\label{AGP}
	\begin{algorithmic}[1]
		\Require A simulator for \eqref{spde_l}--\eqref{spde_b}, the probability density function of $\xi$, and  the number of training points $N_{train}$.
		\State Conduct ANOVA decomposition using Algorithm~\ref{AA} to obtain an effective index set $\pP$
and data sets  
$\Theta_t:=\{y_t^{(k)}=u_t(\xi_t^{(k)})\, |\, \xi_t^{(k)}\in \Xi_t, \textrm{ and } k=1,\ldots,|\Xi_t| \}$ for $t\in\pP$.
               \For{$t\in \pP$}
		\State Build the local GP model $\hat{u}_t(\xi_t)$ using Algorithm~\ref{data_selection}.
		\EndFor
		\State Assemble the ANOVA-GP emulator: $\hat{u}_{\pP}\left(\xi\right):= \sum_{t \in \pP} \hat{u}\left(\xi_t\right)$.
		\Ensure The ANOVA-GP emulator $\hat{u}_{\pP}\left(\xi\right)$.
	\end{algorithmic}
\end{algorithm}

%% file: Numerical_study.tex
In this section, two kinds of model problems are studied to illustrate the effectiveness of our ANOVA-GP strategy:
stochastic diffusion problems in section \ref{result_diff} and stochastic incompressible flow problems in section \ref{result_stokes}. 
For comparison, a direct combination of Gaussian process modeling and PCA is considered, 
which is referred to as the standard Gaussian process (S-GP) in the following.
While S-GP is originally developed by \cite{higdon2008computer} for model calibration, 
we here modify it to build surrogates for  the problem \eqref{spde_l}--\eqref{spde_b}. 
The S-GP emulator for  \eqref{spde_l}--\eqref{spde_b}  is denoted by $u_{SGP}(\xi)$,
and details of our setting for constructing  $u_{SGP}(\xi)$ are summarized in Algorithm \ref{SGP}. 
Using the notation in Algorithm \ref{SGP}, 
the predictive mean of  $u_{SGP}(\xi)$  is denoted by 
\begin{eqnarray}
\overline{u}_{SGP}\left(\xi\right):=V\overline{U}+\mu, \label{sgp-mean}
\end{eqnarray}
where $\overline{U}:=[\mean_1,\ldots,\mean_R]^T$, 
$\mean_1,\ldots,\mean_R$ and $\mu$ are defined in Algorithm \ref{SGP}. 


\begin{algorithm}[]
	\caption{Standard  Gaussian process (S-GP) modeling.}
	\label{SGP}
	\begin{algorithmic}[1]
	       \Require A simulator for \eqref{spde_l}--\eqref{spde_b}, the probability density function of $\xi$ and the number of training points $N$. 
                \State Generate $N$ samples of $\xi$: $\left\{\xi^{(j)}\right\}^N_{j=1}$.
                \State Compute the simulator outputs: $\left\{y^{(j)}=u\left(\xi^{(j)}\right)\right\}^N_{j=1}$.
                 \State Conduct PCA using Algorithm~\ref{PCA} with the  input data set  $\left\{y^{(j)}=u\left(\xi^{(j)}\right)\right\}^N_{j=1}$, 
                 to obtain the sample mean $\mu$, the principle component matrix $V:=[v_1,\ldots,v_R]$, and the data vectors  $\alpha_{r}$ for $r=1,\ldots,R$. 
                 \State Construct GP models for PCA modes: $U=[U_{1},\ldots,U_{R}]^T\in \dsR^R$ with 
		$U_{r}:=\mathcal{GP}\lt(\mean_r(\xi,\beta), \var_r\lt(\xi,\beta\rt)\rt)$ for $r=1,\ldots,R$ (see \eqref{loglike}--\eqref{gp}). 
		\State Construct the S-GP emulator:  $u_{SGP}(\xi)=VU(\xi)+\mu$. 
		\Ensure The S-GP emulator  $u_{SGP}(\xi)$. 
	\end{algorithmic}
\end{algorithm}

\subsection{Test problem 1: diffusion problems}\label{result_diff}
\begin{figure}[htpb!]
	\centering  
	\subfigure[$6\times6=36$ subdomains]{
		\label{Fig1sub2}
		\includegraphics[width=0.45\textwidth]{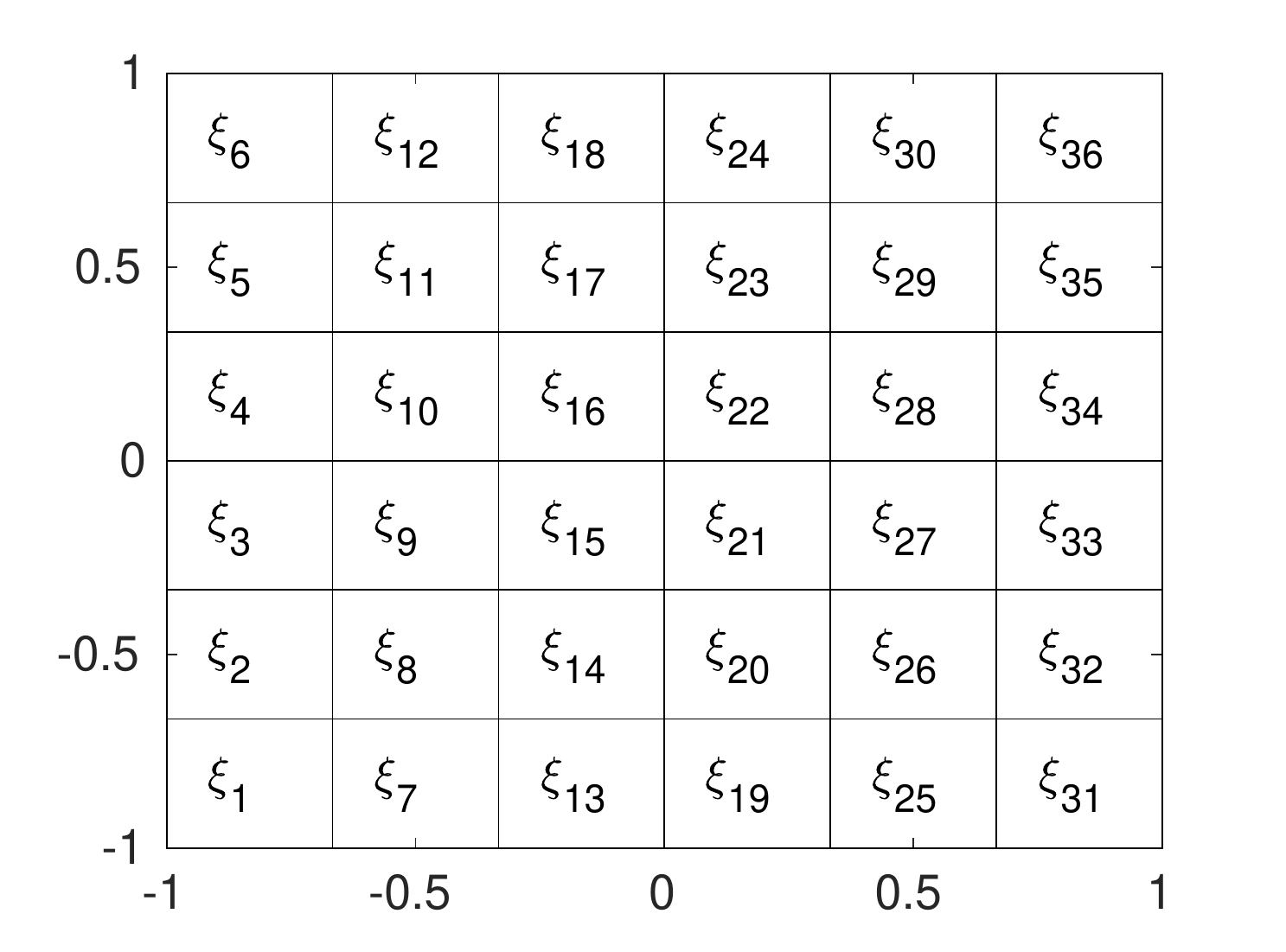}}
	\subfigure[$8\times8=64$ subdomains]{
		\label{Fig1sub3}
		\includegraphics[width=0.45\textwidth]{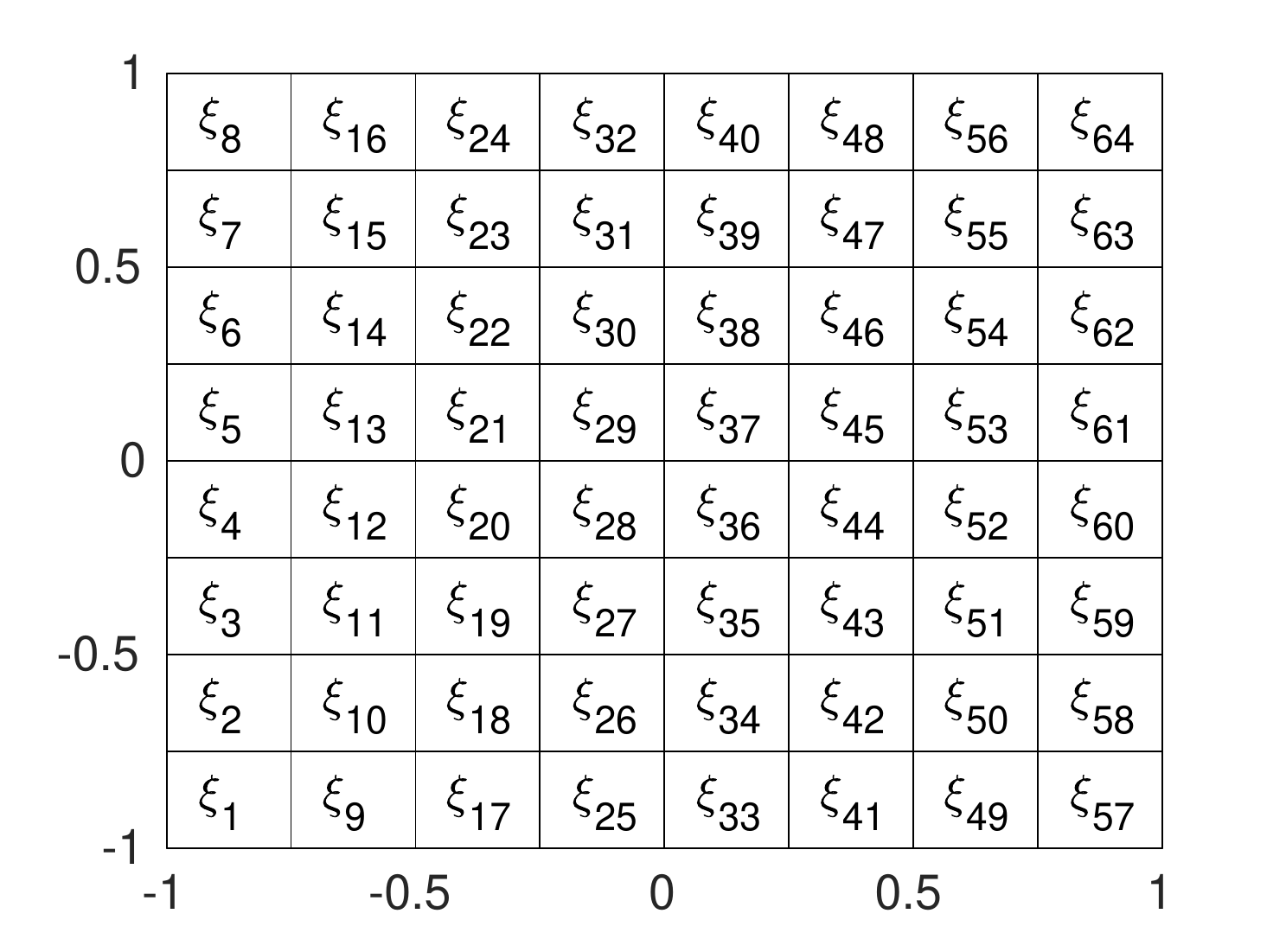}}
	\caption{Partitionings of the physical domain with $N_D = 36$ and $64$ subdomains, test problem 1.}
	\label{diff_domain}
\end{figure}

We consider the following governing equations posed on the physical domain $D\ =\ (-1,\ 1) \times (-1,\ 1)$:
\begin{eqnarray} 
- \nabla \cdot \lt[a\lt(x, \xi\rt) \nabla u_{sol}\lt(x, \xi\rt)\rt] &= 1\quad &  \textrm{in} \quad D  \times \Gamma, \label{diff1}\\
u_{sol}\lt(x, \xi\rt) &= 0\quad & \textrm{on}\quad \partial D  \times \Gamma. \label{diff2}
\end{eqnarray}
Dividing the physical domain $D$ into $N_D$ subdomains, each of which is denoted by $D_k$ for $k=1,\ldots,N_D$,
the permeability coefficient $a(x, \xi)$ is defined to be a piecewise constant function
\begin{equation}
a\lt(x, \xi\rt)|_{D_k} = \xi_k,\ k=1,\dots,N_D, 
\end{equation} 
where $\xi_1,\ldots,\xi_{N_D}$ are independently and uniformly distributed in $[0.01,1]$. 
Two cases of physical domain partitionings are considered, which are shown in Figure \ref{diff_domain} and
include $36$ and $64$ parameters respectively. 
For each realization of $\xi$, the simulator for \eqref{diff1}--\eqref{diff2} is set to the finite element method  \cite{braess2007finite, Elman2014},
where a bilinear $\dsQ{1}$ finite element approximation is used to discretize the physical domain with a $65\times 65$ grid, 
i.e., the dimension of the simulator output is $d=4225$.  

As discussed in section \ref{section_alg}, the first step of our ANOVA-GP strategy is to conduct ANOVA decomposition 
for \eqref{diff1}--\eqref{diff2} using Algorithm \ref{AA}. In Algorithm \ref{AA}, the quadrature rule is set to the tensor products of 
one-dimensional Clenshaw-Curtis quadrature with five quadrature points \cite{kli07}, and the tolerance for selecting effective indices
is set to $tol_{index}=10^{-4}$. The tolerance of PCA (in Algorithm \ref{PCA}) for both ANOVA-GP and S-GP are set to $tol_{pca}=10^{-2}$. 
Table \ref{table_anovaterms} shows sizes of the index sets $\pPp_i$ constructed by \eqref{adaptive} 
and sizes of the selected index sets $\pP_i$ 
at each ANOVA order $i=1,2,3$.
For the two cases of physical domain partitionings ($N_D=36$ and $64$),
all first order indices and a fraction of second order indices are selected, 
while there is no third order index selected, which is consistent with the results in \cite{yang2012adaptive,liao2016reduced}. 

\begin{table}[!htpb]
	\caption{Number of effective ANOVA terms, test problem 1.} 
	\renewcommand\arraystretch{1.5}
	\centering	
	\begin{tabular}{c|cccccc}  
		\hline
		$N_D$ & $|\widehat{\mathcal{J}_1}|$ & $|\mathcal{J}_1|$ & $|\widehat{\mathcal{J}_2}|$ & $|\mathcal{J}_2|$ & $|\widehat{\mathcal{J}_3}|$ & $|\mathcal{J}_3|$ \\ \hline
		$36$ &  36 & 36 & 630 &  100 & 80 & 0   \\
		$64$ &  64 & 64 & 2016 &  172 & 120 & 0  \\
		\hline
	\end{tabular}
	\label{table_anovaterms}
\end{table}

\begin{figure}[htpb!]
	\begin{multicols}{2}
		\centering  
		\subfigure[$N_{agp}=30$, $N_{sgp}=4080$, for $N_D=36$ subdomains]{
			\label{Fig6sub1}
			\includegraphics[width=0.45\textwidth]{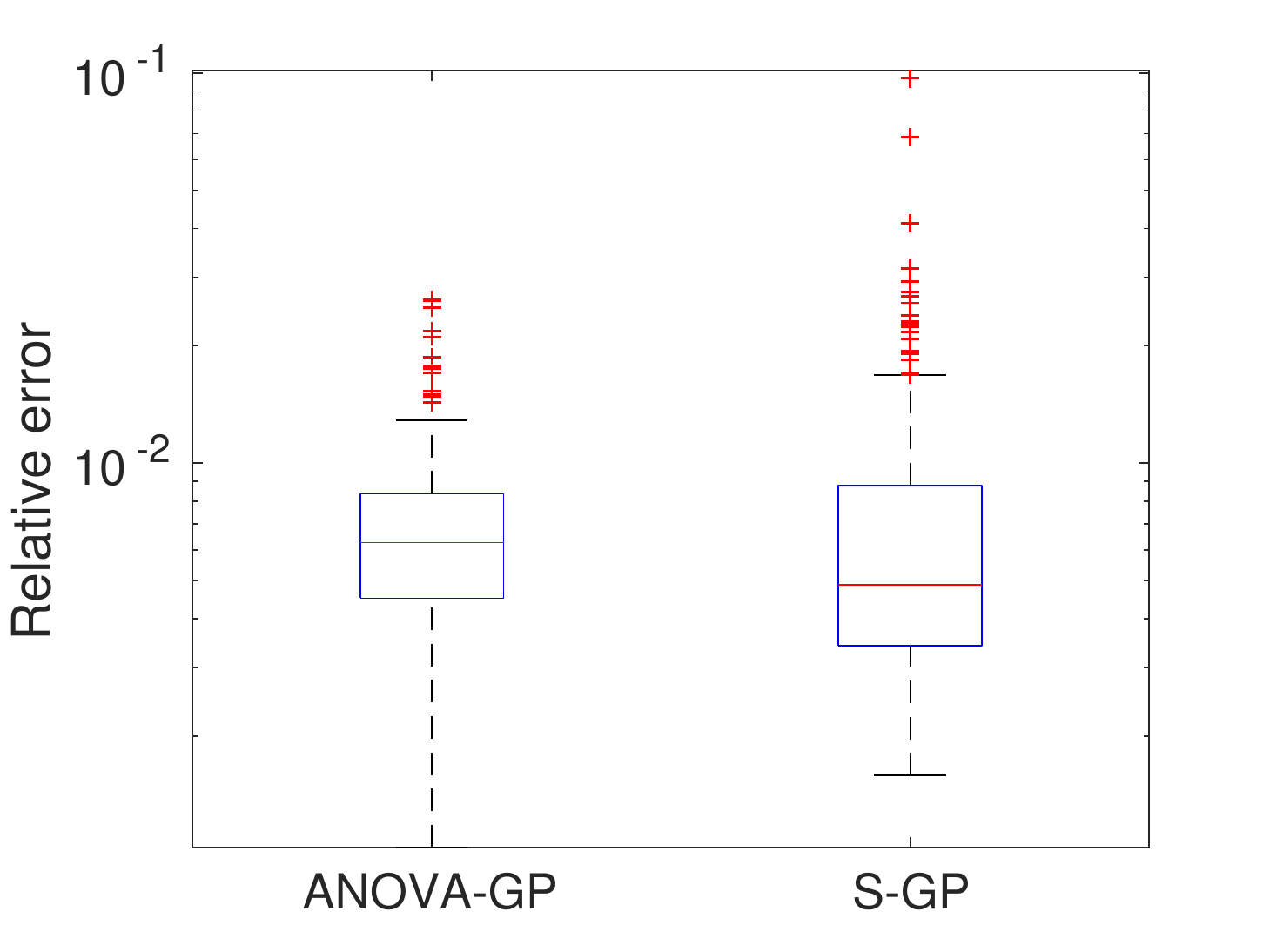}}
		\subfigure[$N_{agp}=50$, $N_{sgp}=6800$, for $N_D=36$ subdomains]{
			\label{Fig8sub1}
			\includegraphics[width=0.45\textwidth]{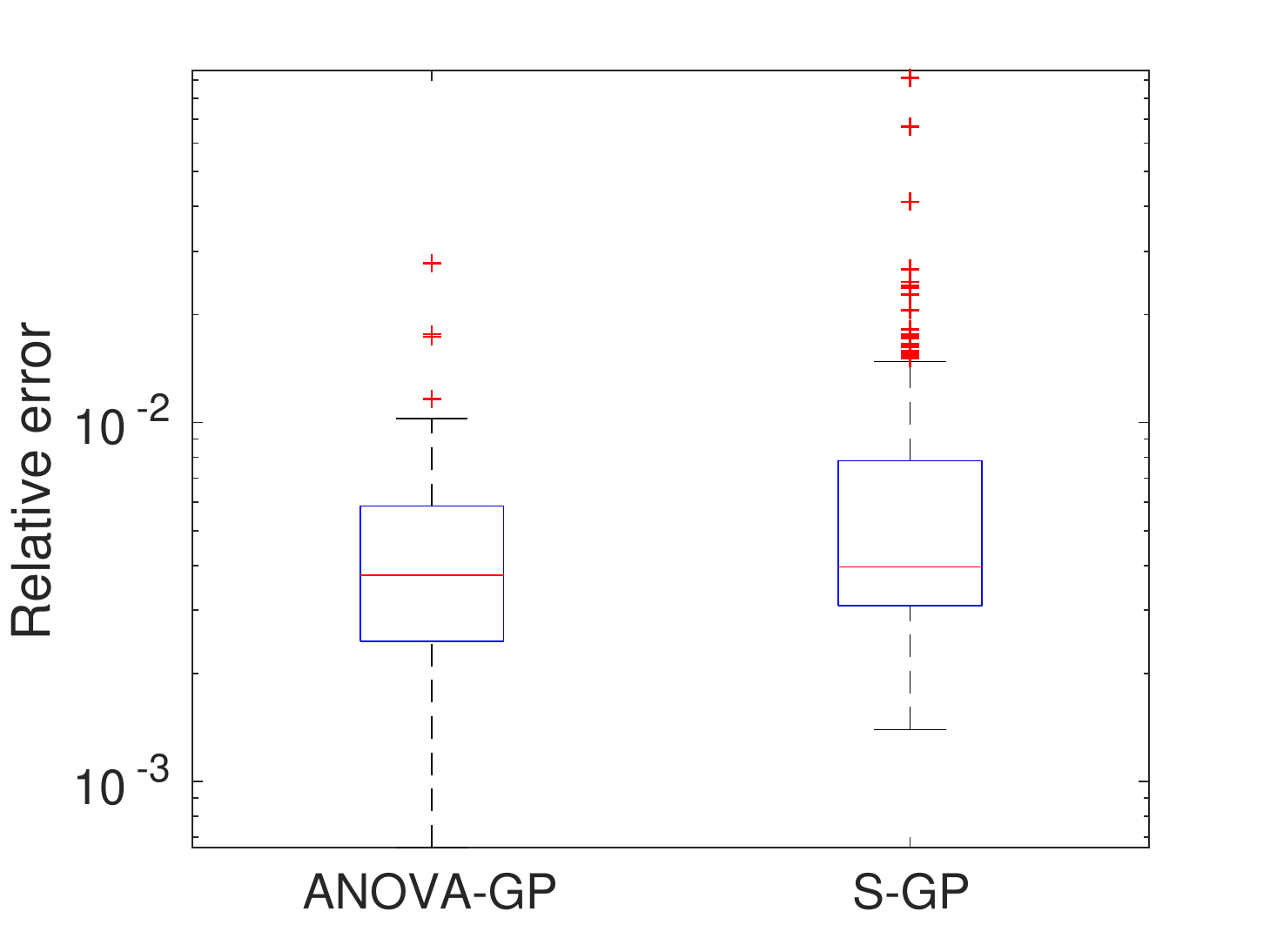}}
		\subfigure[$N_{agp}=70$, $N_{sgp}=9520$, for $N_D=36$ subdomains]{
			\label{Fig9sub1}
			\includegraphics[width=0.45\textwidth]{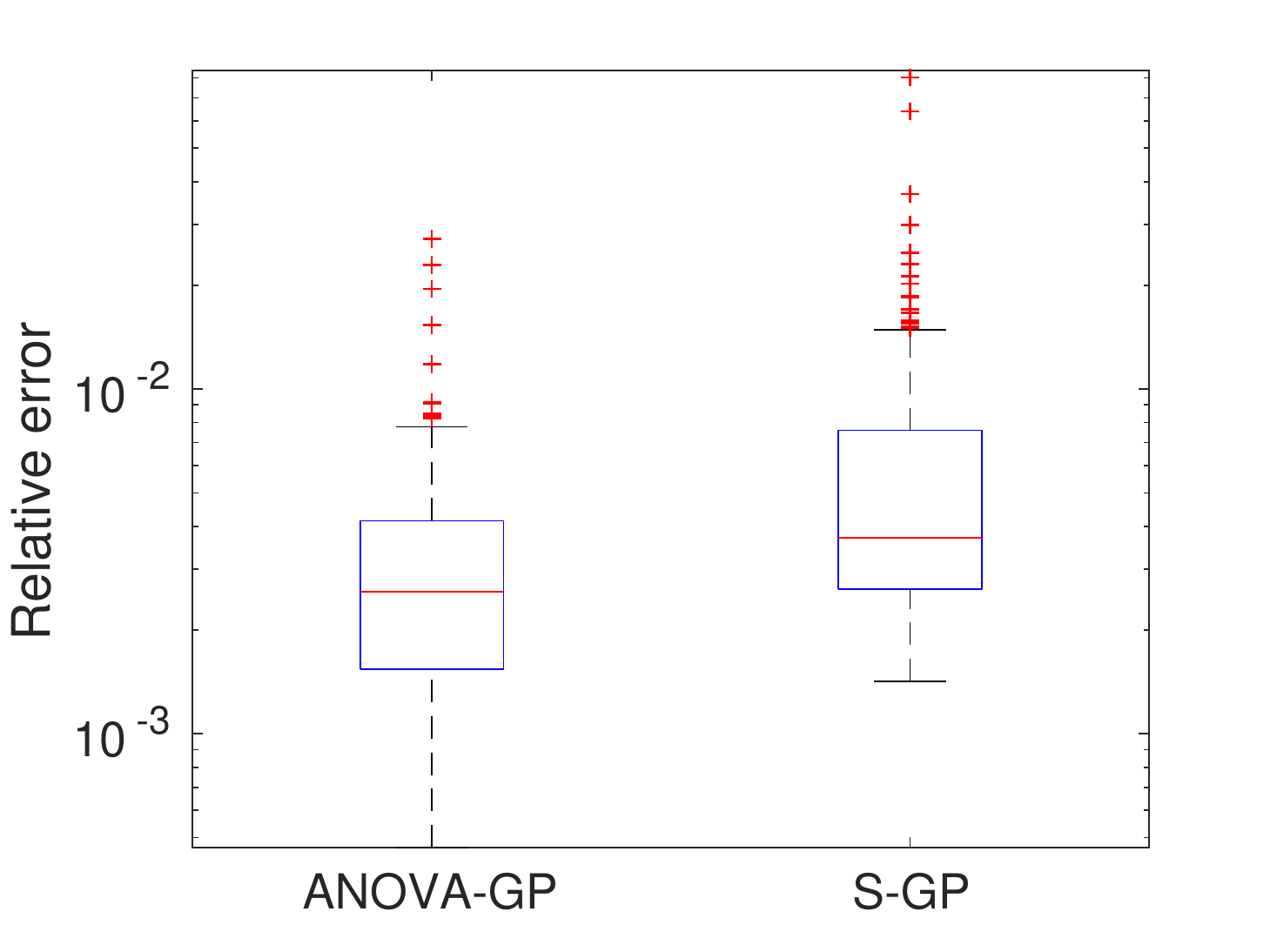}}
		\subfigure[$N_{agp}=30$, $N_{sgp}=7080$, for $N_D=64$ subdomains]{
			\label{Fig6sub2}
			\includegraphics[width=0.45\textwidth]{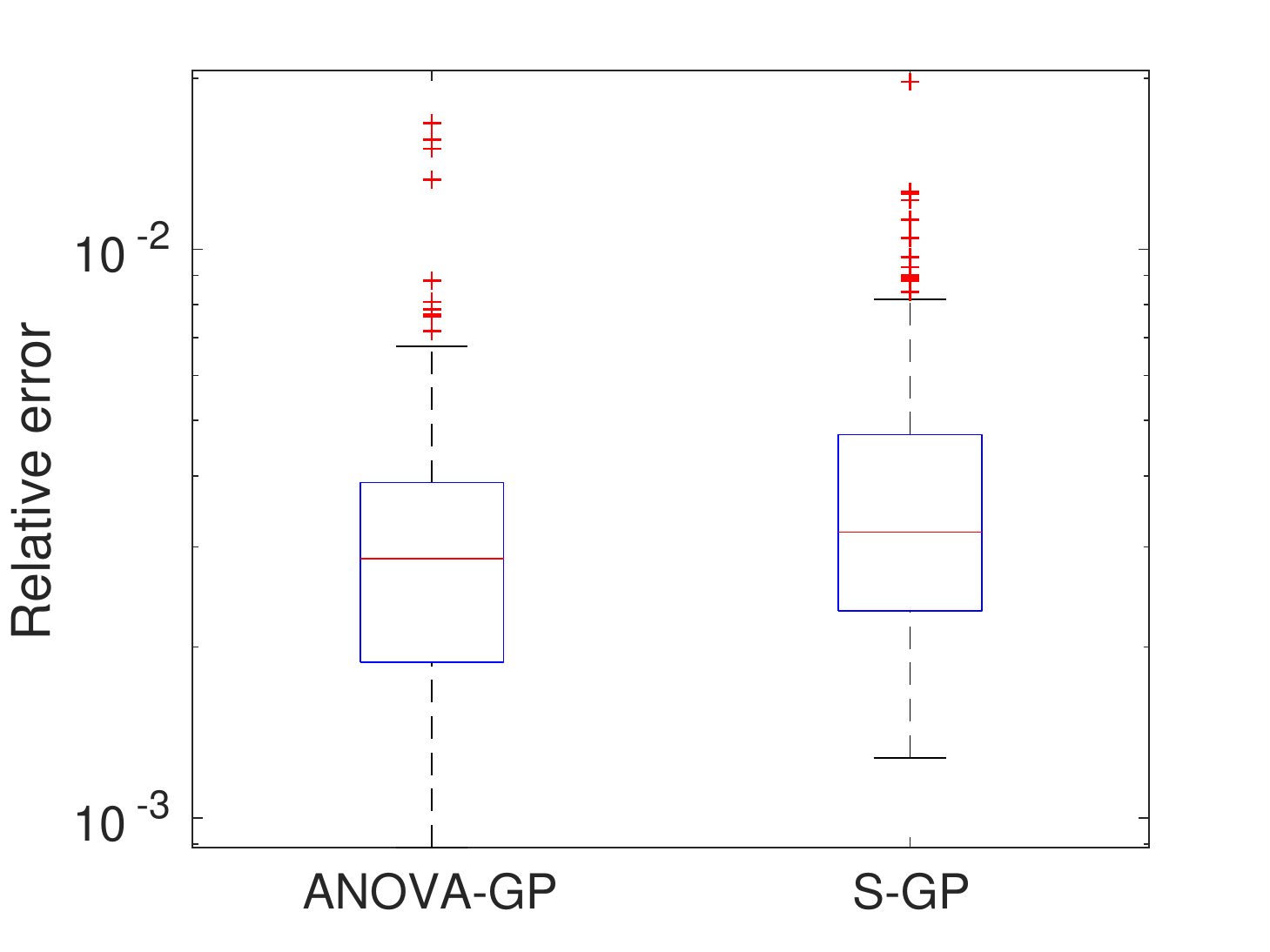}}
		\subfigure[$N_{agp}=50$, $N_{sgp}=11800$, for $N_D=64$ subdomains]{
			\label{Fig8sub2}
			\includegraphics[width=0.45\textwidth]{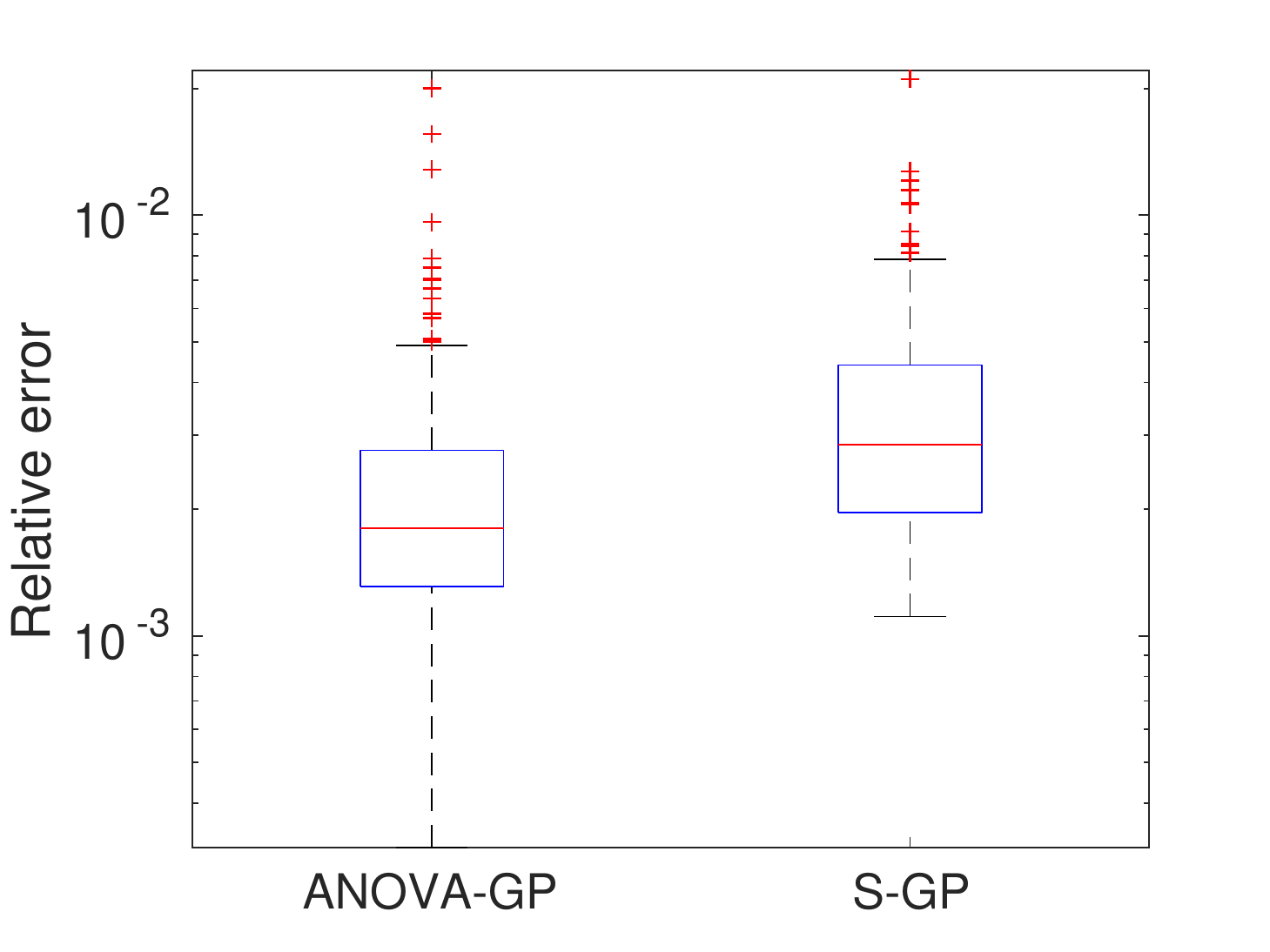}}
		\subfigure[$N_{agp}=70$, $N_{sgp}=16520$, for $N_D=64$ subdomains]{
			\label{Fig9sub2}
			\includegraphics[width=0.45\textwidth]{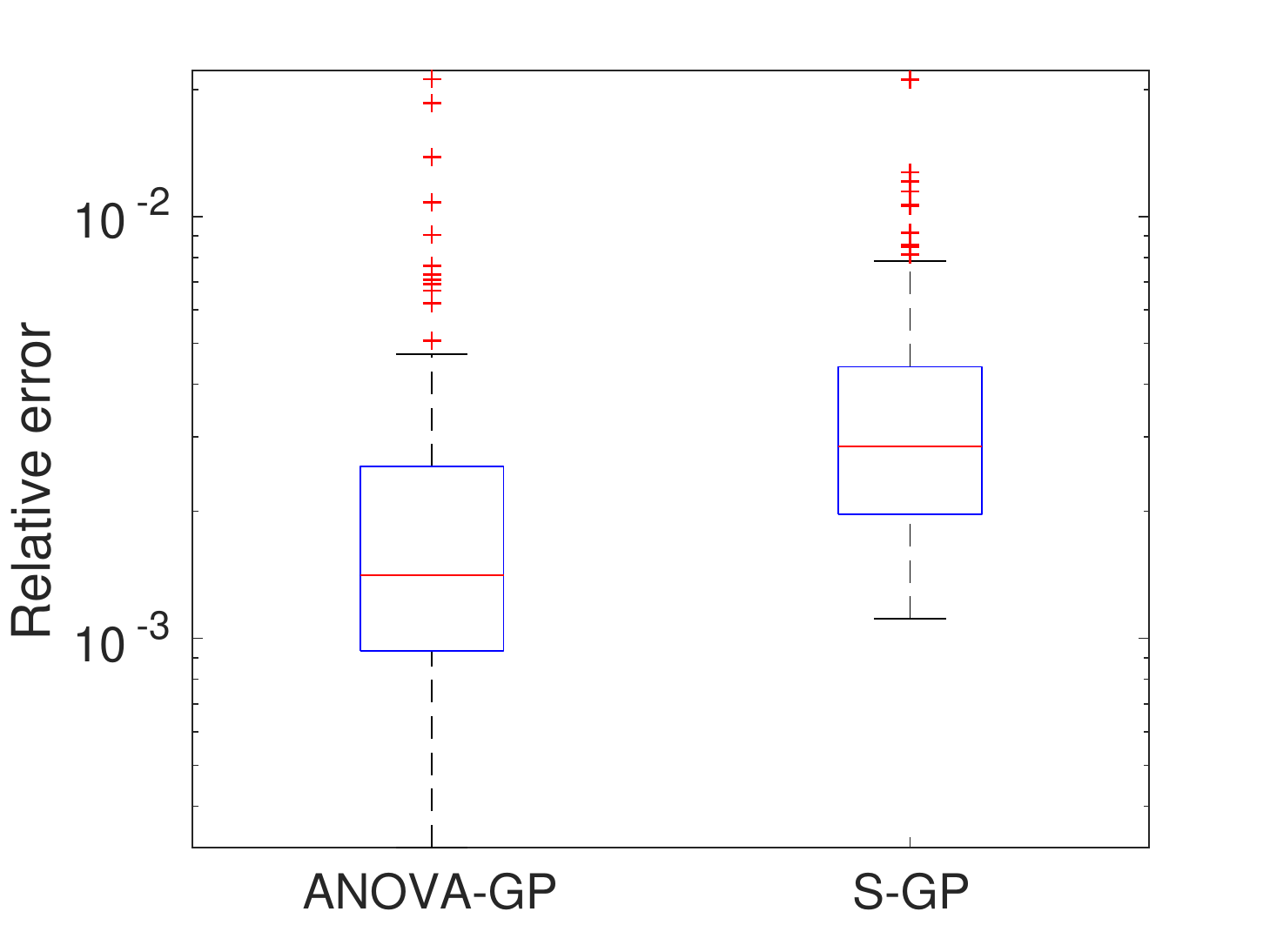}}
	\end{multicols}
	\caption{Relative errors of ANOVA-GP and S-GP for $200$ test data points, test problem 1.}
	\label{Box_pc_16}
\end{figure}

Accuracy of our ANOVA-GP emulator and the standard GP (S-GP) emulator is assessed as follows.
First, $200$ samples of $\xi$ is generated and denoted by $\{\xi^{(j)}\}^{200}_{j=1}$, and 
the corresponding simulator output is computed and denoted by $\{y^{(j)}=u(\xi^{(j)})\}^{200}_{j=1}$.  
Next, for each $j=1,\ldots,200$, we consider the relative error used in \cite{xing2016manifold}
\begin{equation}
\mathrm{Relative\ error} = \frac{\left\Vert y_p^{(j)} - y^{(j)} \right\Vert^2}{\left\Vert y^{(j)} \right\Vert^2}, \label{r_error}
\end{equation} 
 where $\Vert \cdot \Vert$ is  the standard Euclidean norm.
In \eqref{r_error}, 
$y^{(j)}_p$ refers to the predictive mean $ \overline{u}_{\pP}(\xi^{(j)})$ in \eqref{anova-gp-mean} when assessing the errors of ANOVA-GP, and 
it refers to $\overline{u}_{SGP}(\xi^{(j)})$ in  \eqref{sgp-mean} when assessing the errors of S-GP.
Different numbers of training data points to build the GP models are tested. 
For ANOVA-GP (Algorithm \ref{AGP}),  the following numbers of training points for each 
ANOVA index $t\in \pP$ are tested: $N_{train}=30$, $50$ and $70$. 
For a fair comparison, for S-GP (Algorithm \ref{SGP}),  the number of training points $N$ is set to the number of all training points 
generated in Algorithm \ref{AGP}, which is $N=N_{train}\times (|\pP|-1)$, where $|\pP|-1$ is the number of ANOVA terms except the zeroth order term
(the zeroth order term $u(c)$ is given).  In the following, we denote $N_{sgp}=N$ for S-GP,
and $N_{agp}=N_{train}$ for ANOVA-GP.
Figure \ref{Box_pc_16} shows the errors for the test problem with $32$ and $64$ subdomains,
and it is clear that as the number of training points increases, our ANOVA-GP has smaller errors than S-GP. 

\begin{figure}[htpb!]
	\begin{multicols}{2}
		\centering  	
		\subfigure[Simulator output, $N_D=36$]{
			\label{Fig7sub4}
			\includegraphics[width=0.45\textwidth]{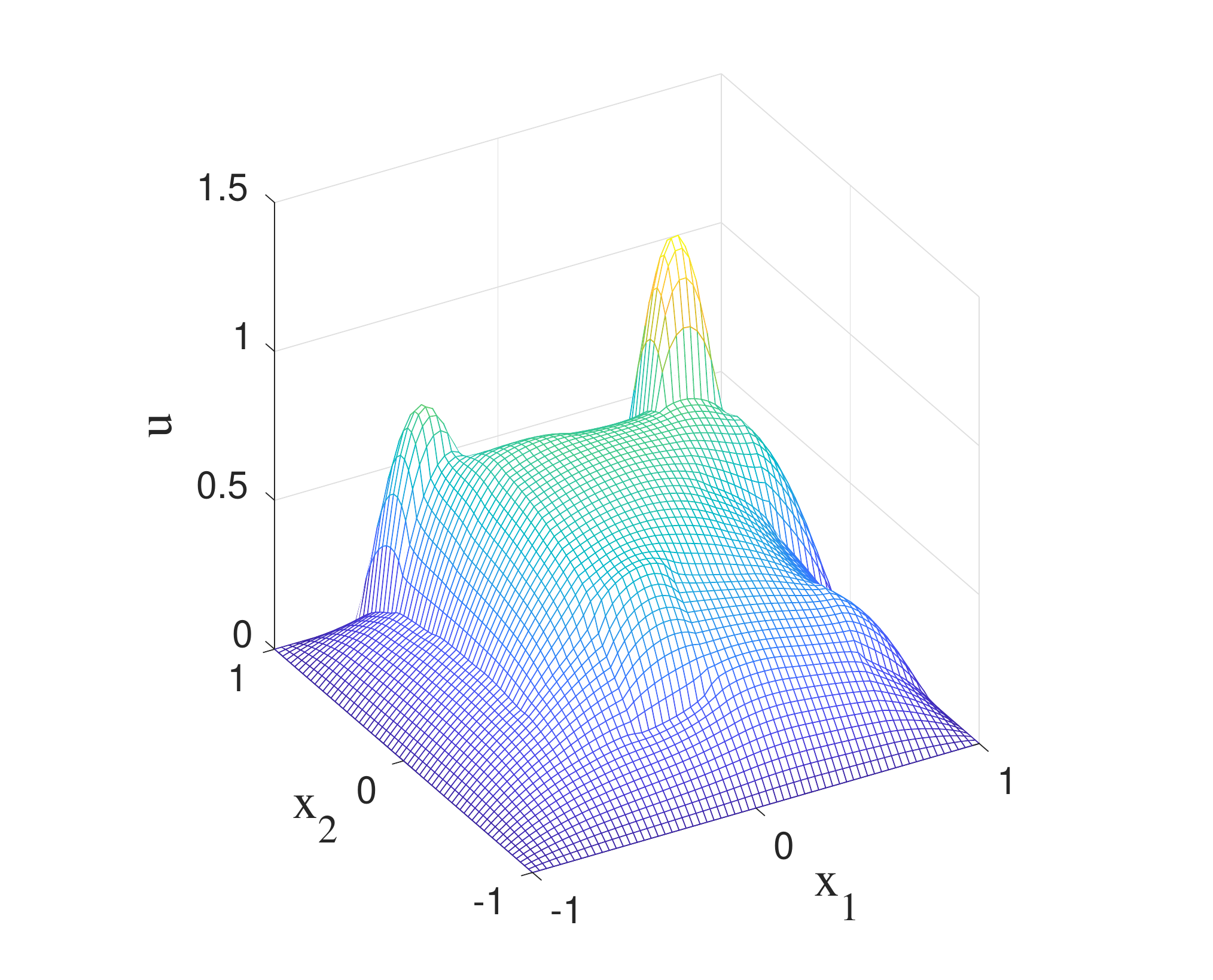}}
		\subfigure[Predictive mean of S-GP, $N_{sgp}=9520$, $N_D=36$]{
			\label{Fig7sub6}
			\includegraphics[width=0.45\textwidth]{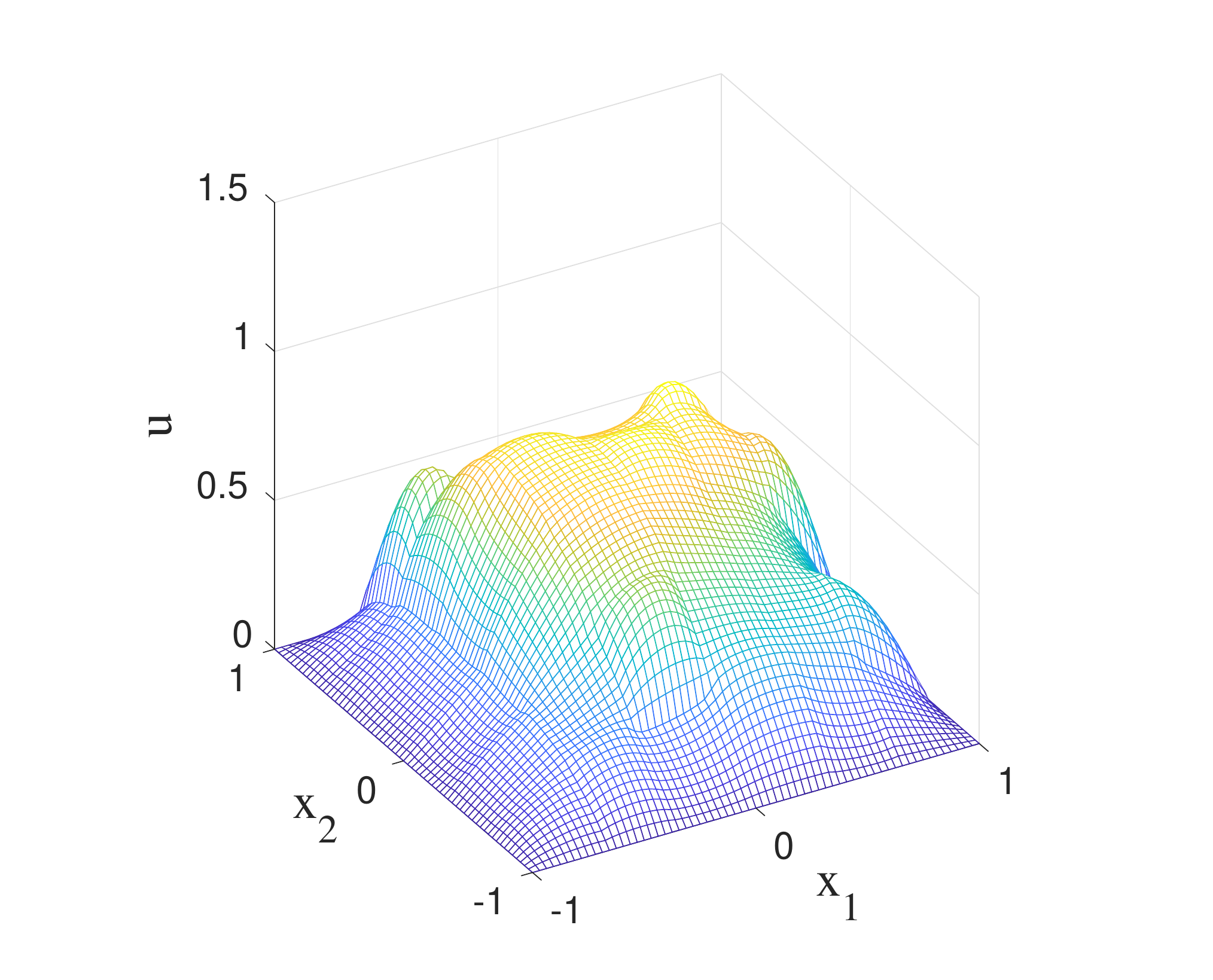}}
		\subfigure[Predictive mean of ANOVA-GP, $N_{agp}=70$, $N_D=36$]{
			\label{Fig7sub8}
			\includegraphics[width=0.45\textwidth]{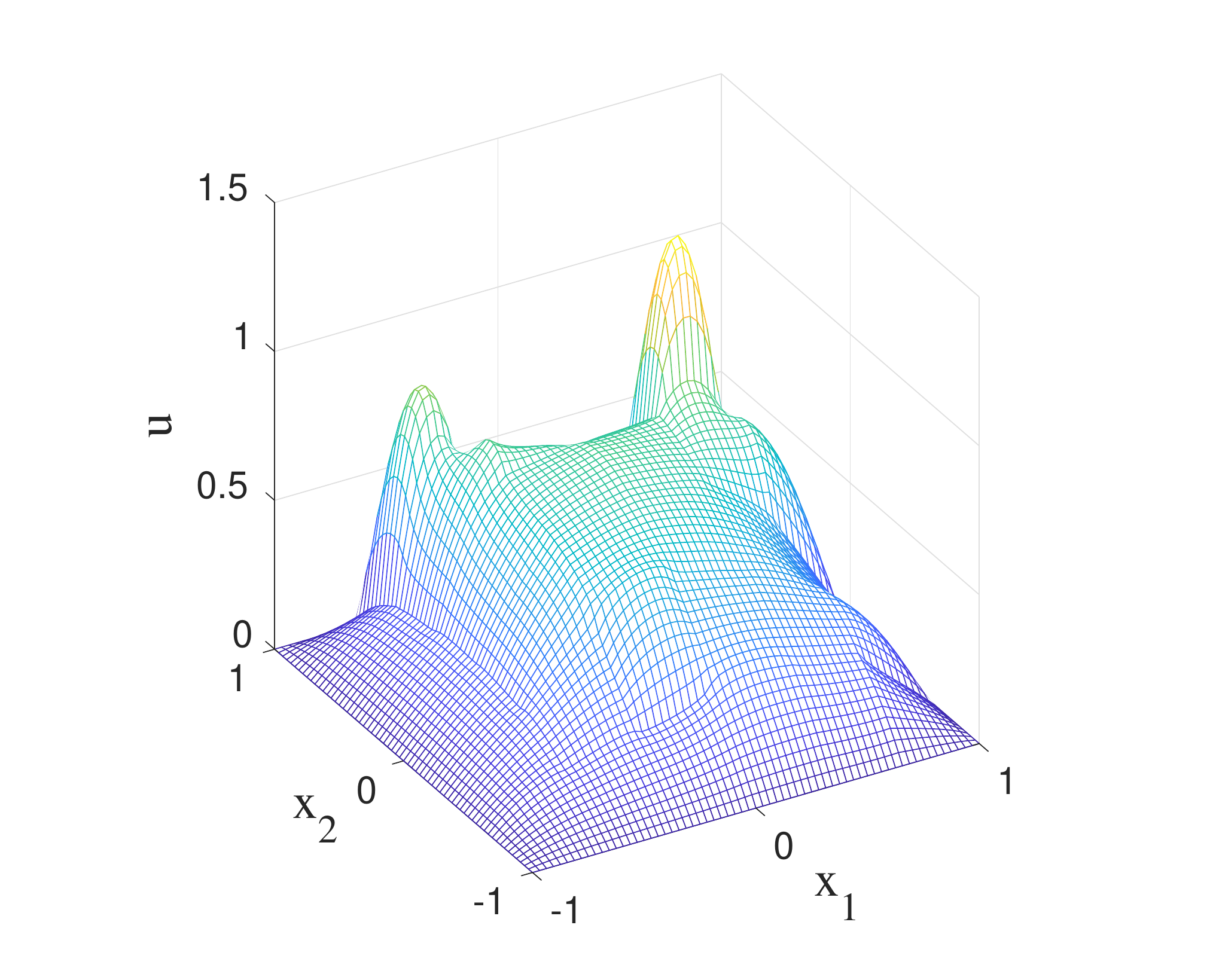}}
		\subfigure[Simulator output, $N_D=64$]{
			\label{Fig7sub5}
			\includegraphics[width=0.45\textwidth]{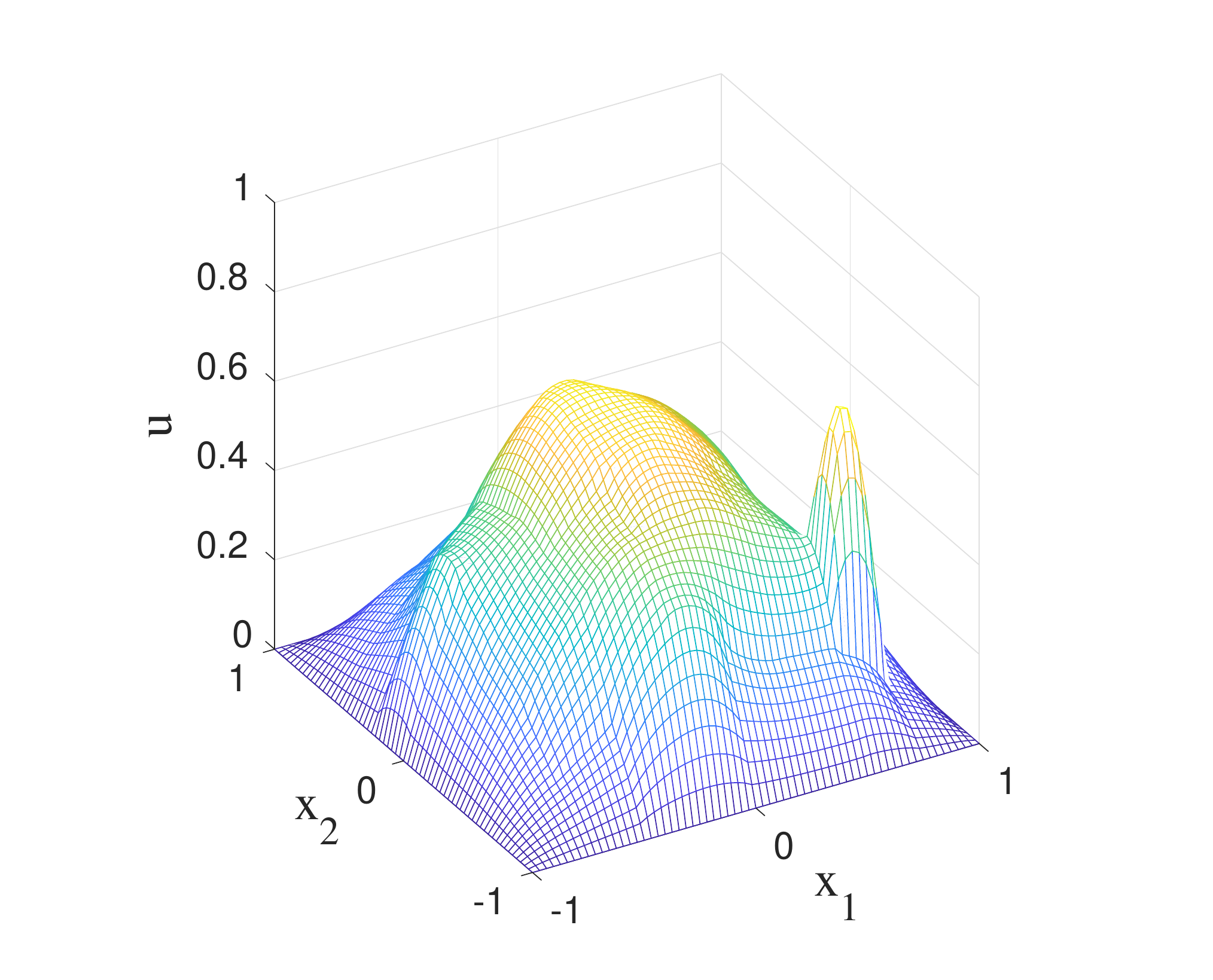}}
		\subfigure[Predictive mean of S-GP, $N_{sgp}=16520$, $N_D=64$]{
			\label{Fig7sub7}
			\includegraphics[width=0.45\textwidth]{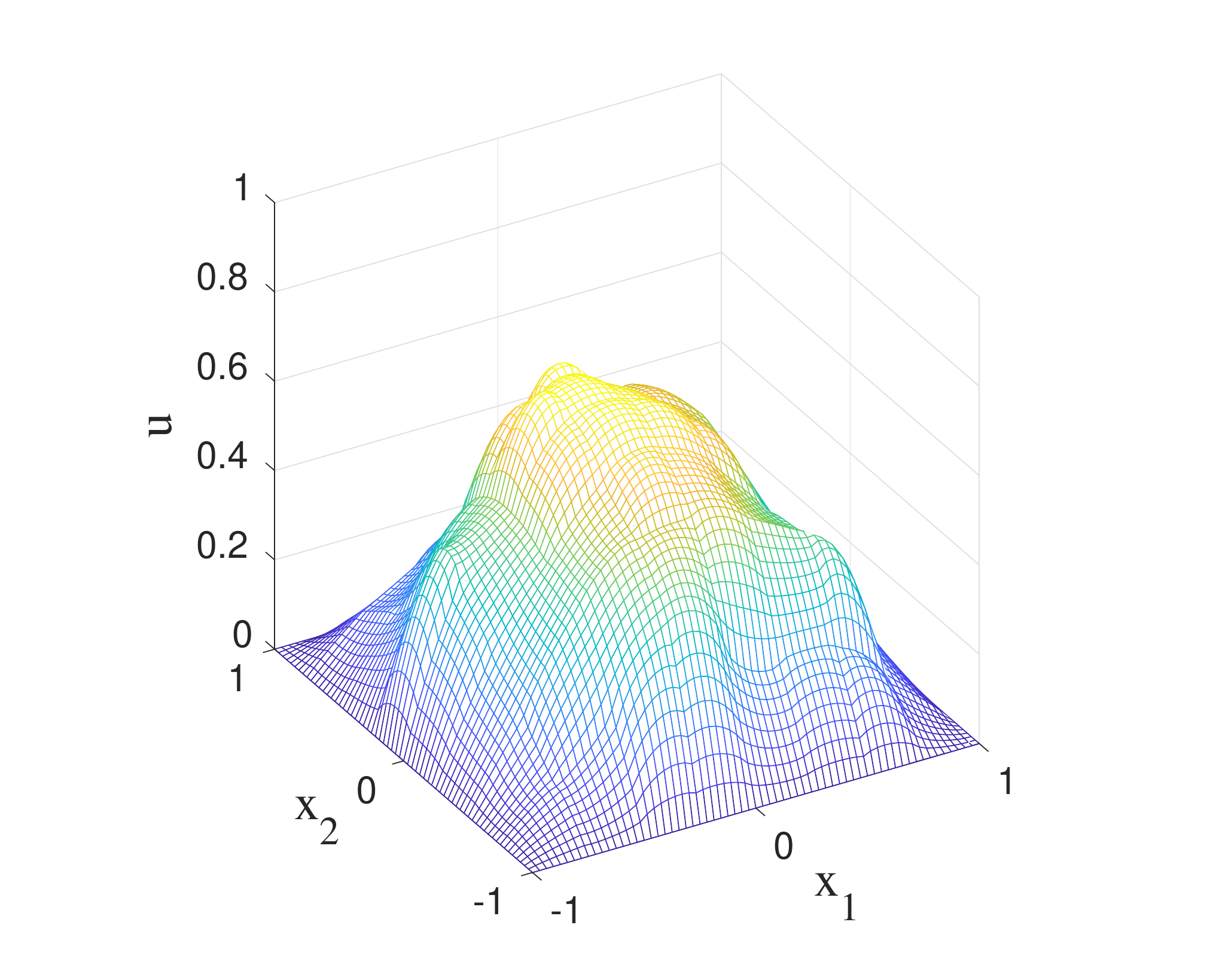}}
		\subfigure[Predictive mean of ANOVA-GP, $N_{agp}=70$, $N_D=64$]{
			\label{Fig7sub9}
			\includegraphics[width=0.45\textwidth]{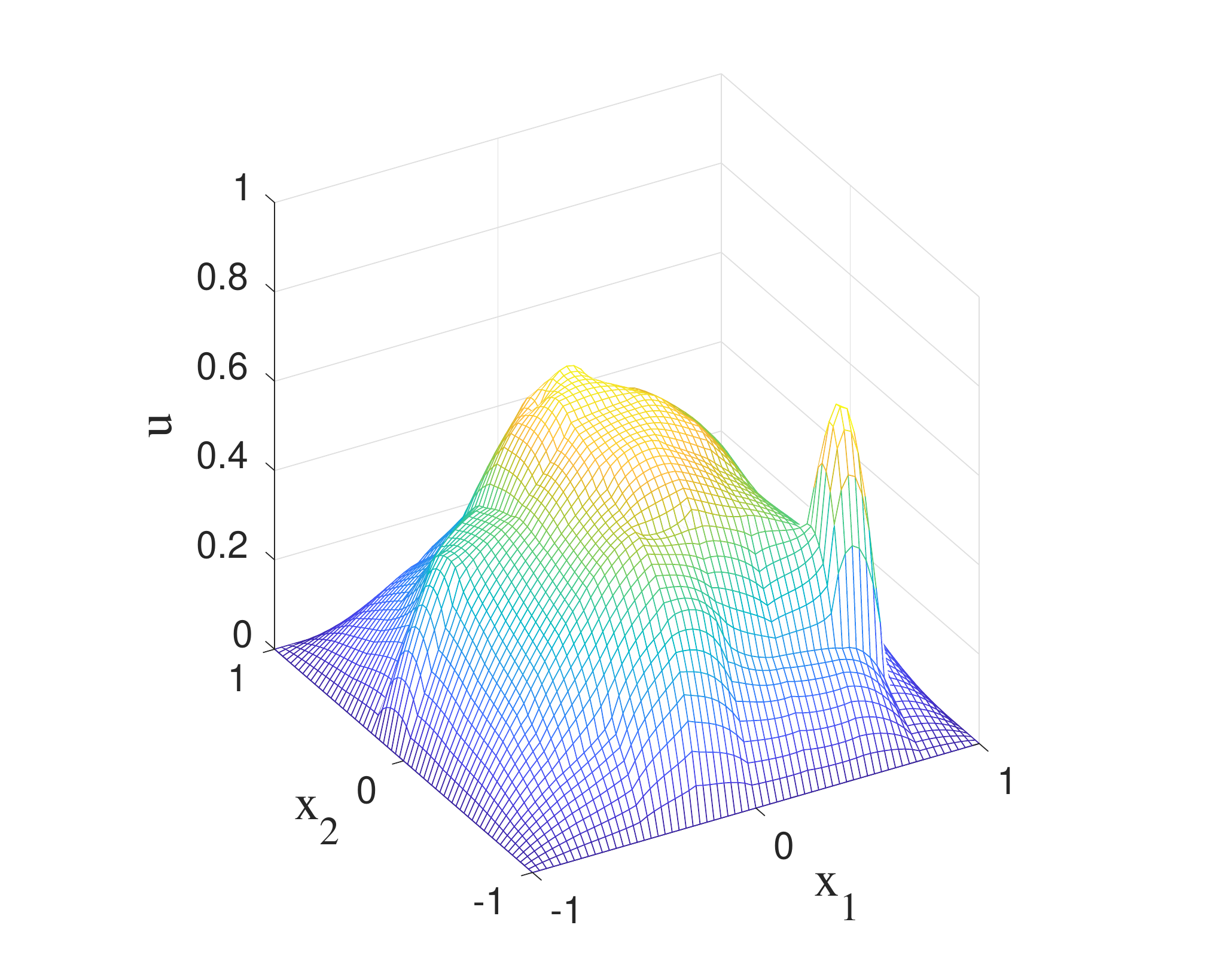}}
	\end{multicols}
	\caption{Examples of predictions made by S-GP and ANOVA-GP, and the simulator outputs
	for both $N_D=36$ and $64$ subdomains, test problem 1.}
	\label{fig:pc_realization}
\end{figure}

Figure \ref{fig:pc_realization} shows the simulator output and the emulator predictive means   
corresponding to a given realization of $\xi$. It is clear that the predictive means of ANOVA-GP are much more accurate than 
 those of S-GP. For example, 
looking at the simulator output  for the case $N_D=36$ in Figure \ref{fig:pc_realization}(a), 
there is a bump near the top right corner $(1,1)$.
The predictive mean of S-GP in  Figure \ref{fig:pc_realization}(b) is too smooth and 
can not show the bump, while our ANOVA-GP output in Figure \ref{fig:pc_realization}(c) 
can capture all details of the simulator output.
For the case $N_D=64$, the  predictive mean of ANOVA-GP  in  Figure \ref{fig:pc_realization}(f) 
is very close to the simulator output  (shown in Figure \ref{fig:pc_realization}(d)), 
while the predictive mean of S-GP (Figure \ref{fig:pc_realization}(e)) can not capture the bump near $(1,-1)$ which can clearly 
be seen in Figure \ref{fig:pc_realization}(d) and Figure \ref{fig:pc_realization}(f). 

\begin{figure}[htpb!]
	\centering  
	\subfigure[$N_D=36$]{
		\label{Fig2sub2}
		\includegraphics[width=0.45\textwidth]{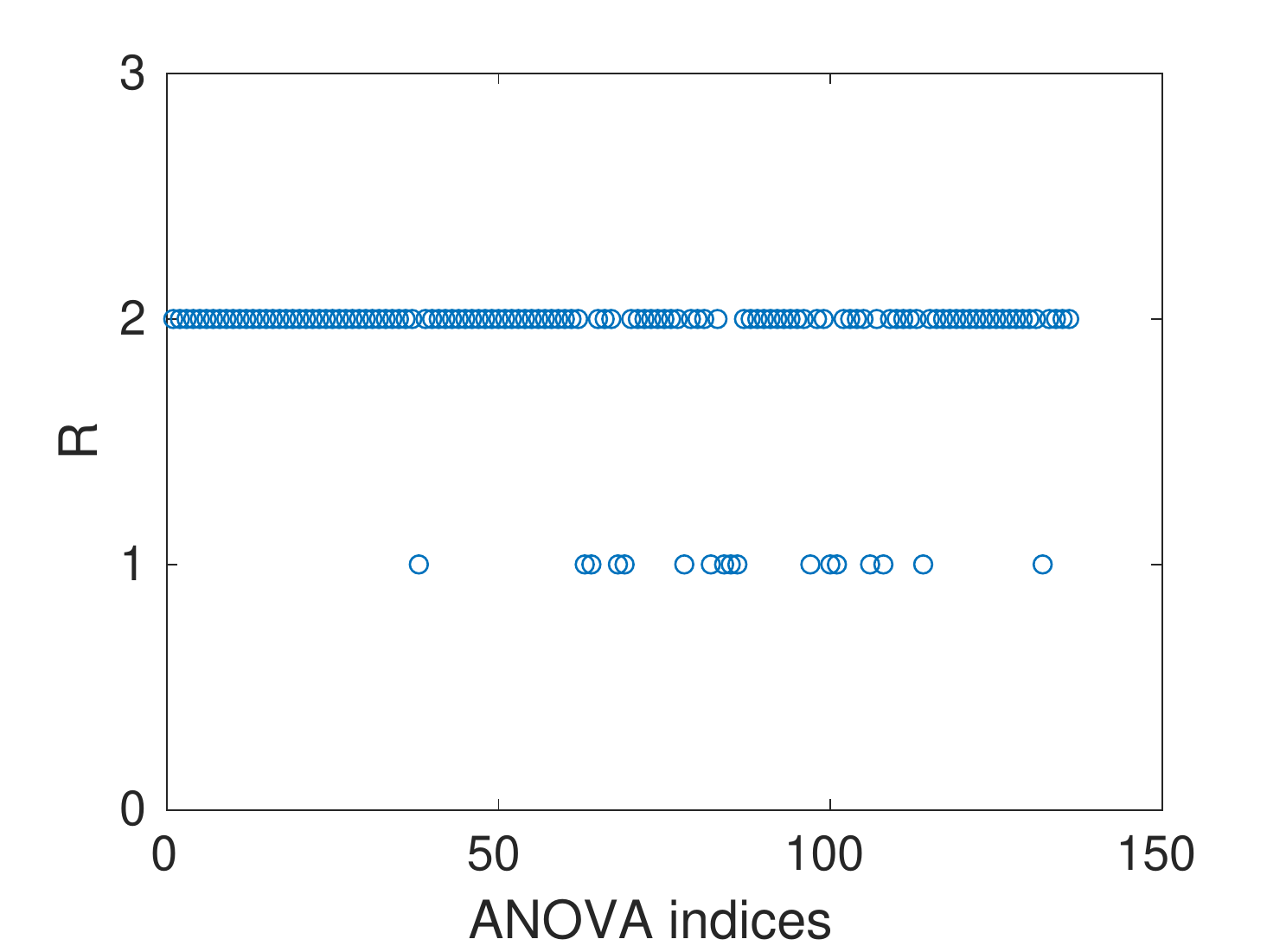}}
	\subfigure[$N_D=64$]{
		\label{Fig2sub3}
		\includegraphics[width=0.45\textwidth]{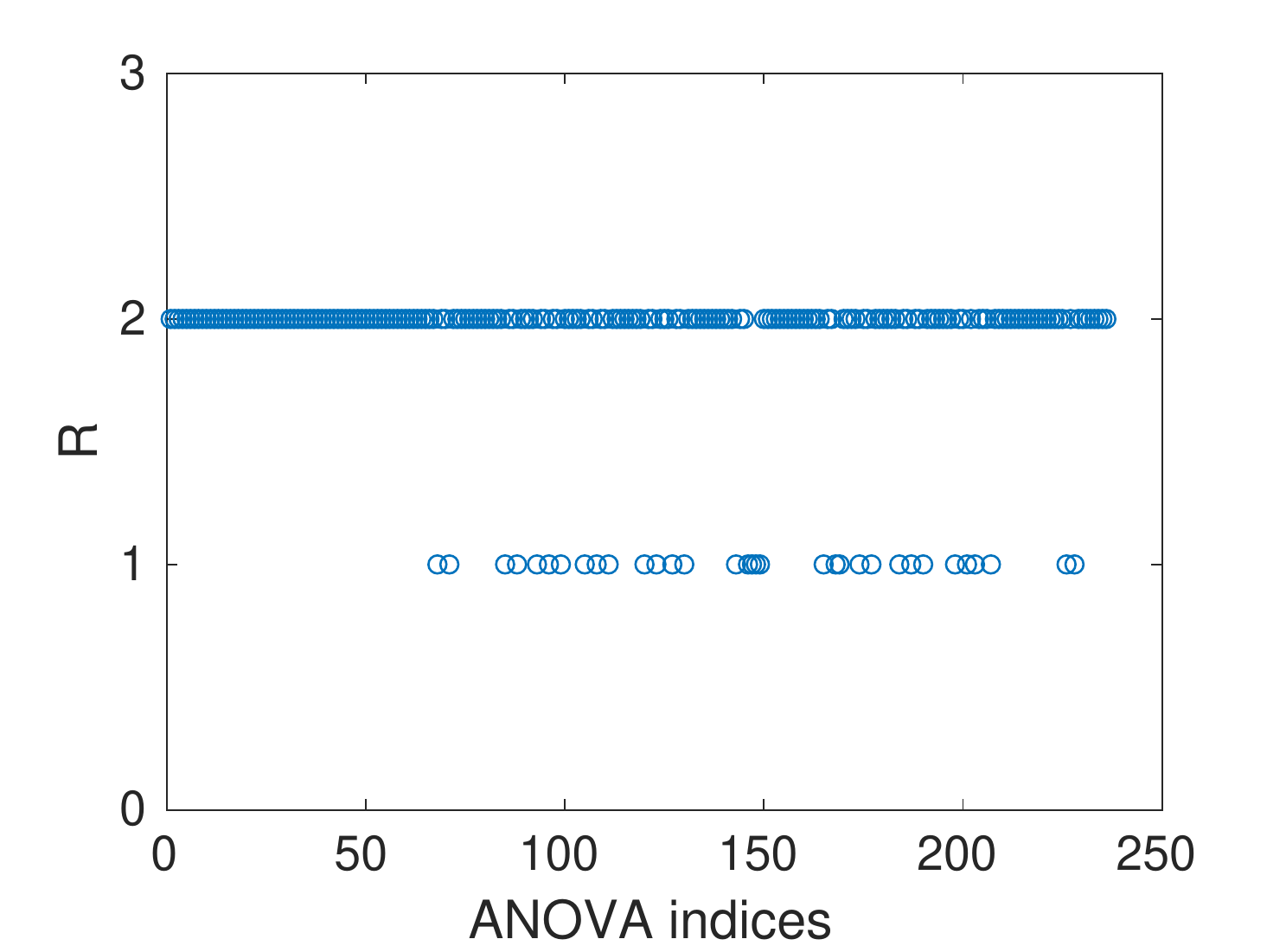}}
	\caption{Number of PCA modes retained, test problem 1.}
	\label{fig:pca_diff}
\end{figure}

For both ANOVA-GP and S-GP, PCA is conducted to result in a reduced dimensional representation of the outputs.
Here, we show results of the case $N_D=36$ with $N_{agp}=70$ for ANOVA-GP and $N_{sgp}=9520$ for S-GP,
and the case $N_D=64$ with $N_{agp}=70$ and $N_{sgp}=16520$. 
For S-GP, the number of PCA modes retained is {60} for the case $N_D=36$, and that is $100$  
for $N_D=64$. 
Figure \ref{fig:pca_diff} shows the number of PCA modes retained for each ANOVA term in ANOVA-GP. 
It is clear that, the numbers are very small---there are at most two PCA modes 
retained for both cases ($N_D=36$ and $N_D=64$). 
In Figure \ref{fig:pca_diff} the ANOVA indices are ordered alphabetically as:
for  any two different indices $t^{(j)}$ and 
$t^{(k)}$ belonging $\pP$, $t^{(j)}$ is ordered before $t^{(k)}$ (i.e., $j<k$), if one of the
following two cases is true: (a)  $|t^{(j)}|<|t^{(k)}|$; (b) $|t^{(j)}|=|t^{(k)}|$ and for the smallest number 
$n\in \{1,\ldots,|t^{(j)}|\}$ such that $t^{(j)}_n\neq t^{(k)}_n$, we have $t^{(j)}_n<t^{(k)}_n$ (where 
 $t^{(j)}_n$ and $t^{(k)}_n$ are the $n$-th components of $t^{(j)}$ and $t^{(k)}$).
So, each local GP model (see line 4 of Algorithm \ref{data_selection}) in ANOVA-GP only involves  a very small number of 
independent GP models, so that training local GP models and using them to conduct predictions are both cheap.

\subsection{Test problem 2: the Stokes problems}\label{result_stokes}
The Stokes equations for this test problem are 
\begin{eqnarray} 
\nabla \cdot \lt[a\lt(x, \xi\rt) \nabla u_{sol}\lt(x, \xi\rt)\rt] + \nabla p_{sol}\lt(x, \xi\rt) &= 0 \quad \mathrm{in} \quad D  \times \Gamma, \label{stokes_1}\\
\div u_{sol}\lt(x, \xi\rt) &= 0 \quad \mathrm{in} \quad D \times \Gamma,\label{stokes_2} \\ 
u_{sol}\lt(x, \xi\rt) &= g \quad \mathrm{on} \quad \partial D \times \Gamma,\label{stokes_3}
\end{eqnarray}
where $D\in \dsR^2$, and 
$u_{sol}(x, \xi) = [u_{sol,1}(x, \xi), u_{sol,2}(x, \xi)]^T$ and  $p_{sol}(x, \xi)$ are  
the flow velocity and the scalar pressure respectively. 
In \eqref{stokes_1}, 
we focus on the situation that there exists uncertainties in the flow viscosity $a(x,\xi)$, which is assumed to be 
a random field with mean function $a_0(\bx)=1$, standard deviation $\sigma=0.5$ and covariance function $Cov(x,x')$
\begin{equation} 
Cov\lt(x,x'\rt) = \sigma^2 \mathrm{exp} \left( -\frac{\vert x_1 - x'_1\vert}{l_c} - \frac{\vert x_2 - x'_2\vert}{l_c} \right). \label{eq_kl_covarfun}
\end{equation}
In \eqref{eq_kl_covarfun}, $x=[x_1,x_2]^T, x'=[x'_1,x'_2]^T\in D$, 
and the correlation length is set to $l_c=0.5$. 
To approximate the random field $a(x, \xi)$, the truncated KL expansion can be applied~\cite{brown1960mean,ghanem2003sfem,schwab2006karhunen}
\begin{equation*}
a\lt(x, \xi\rt) \approx a_0\lt(x\rt) +  \sum\limits_{i=1}^m {\sqrt{\zeta_i} a_i\lt(x\rt) \xi_i},
\end{equation*}
where $\zeta_1\geq,\ldots,\geq \zeta_m$ and $a_1(x),\ldots,a_m(x)$ are the eigenvalues and eigenfunctions of the covariance function~(\ref{eq_kl_covarfun}), $m$ is the number of KL modes retained,
and $\xi = [\xi_1,\dots,\xi_m]^T$ are
uncorrelated random variables. In this test problem, 
$\xi_1,\dots,\xi_m$ are assumed to be independent uniform distributions in $[-1, 1]$. 
We consider the driven cavity flow problem posed on the physical domain $D = (0,1) \times (0,1)$. 
The velocity profile $u = [1, 0]^T$ is imposed on the top boundary $\{[x_1,x_2]^T\ |\ x_1\in(0,1), x_2=1\}$, 
and  the no-slip and no-penetration condition $u = [0, 0]^T$ is applied on all other boundaries. 
The error of the truncated KL expansion depends on the amount of total variance
captured, and we set $m=109$ to capture $95\%$ of the total variance, i.e., ${\sum_{j=1}^{m}\zeta_j}/{(|D|\sigma^2)} > 0.95$,
where $|D|$ refers to the area of $D$ \cite{ghanem2003sfem,powelm09}.  

The simulator for this test problem is set to the  $\bm{Q}_2-\bm{P}_{-1}$ mixed finite element method (biquadratic velocity--linear discontinuous pressure) 
implemented in IFISS~\cite{elman2014ifiss,Elman2014}, 
with the physical domain discretized on a uniform $33 \times 33$ grid, which gives the velocity degrees 
of freedom $N_{u} = 2178$ and 
the pressure degrees of freedom $N_{p} = 768$. 
For each realization of $\xi$,  the simulator output $y$ is defined to be the vector collecting the coefficients of the $\bm{Q}_2-\bm{P}_{-1}$ 
approximation solution for \eqref{stokes_1}--\eqref{stokes_3}, and the dimension of the simulator output is $2946$.

\begin{figure}[htpb!]
	\centering  	
	\includegraphics[width=0.6\linewidth]{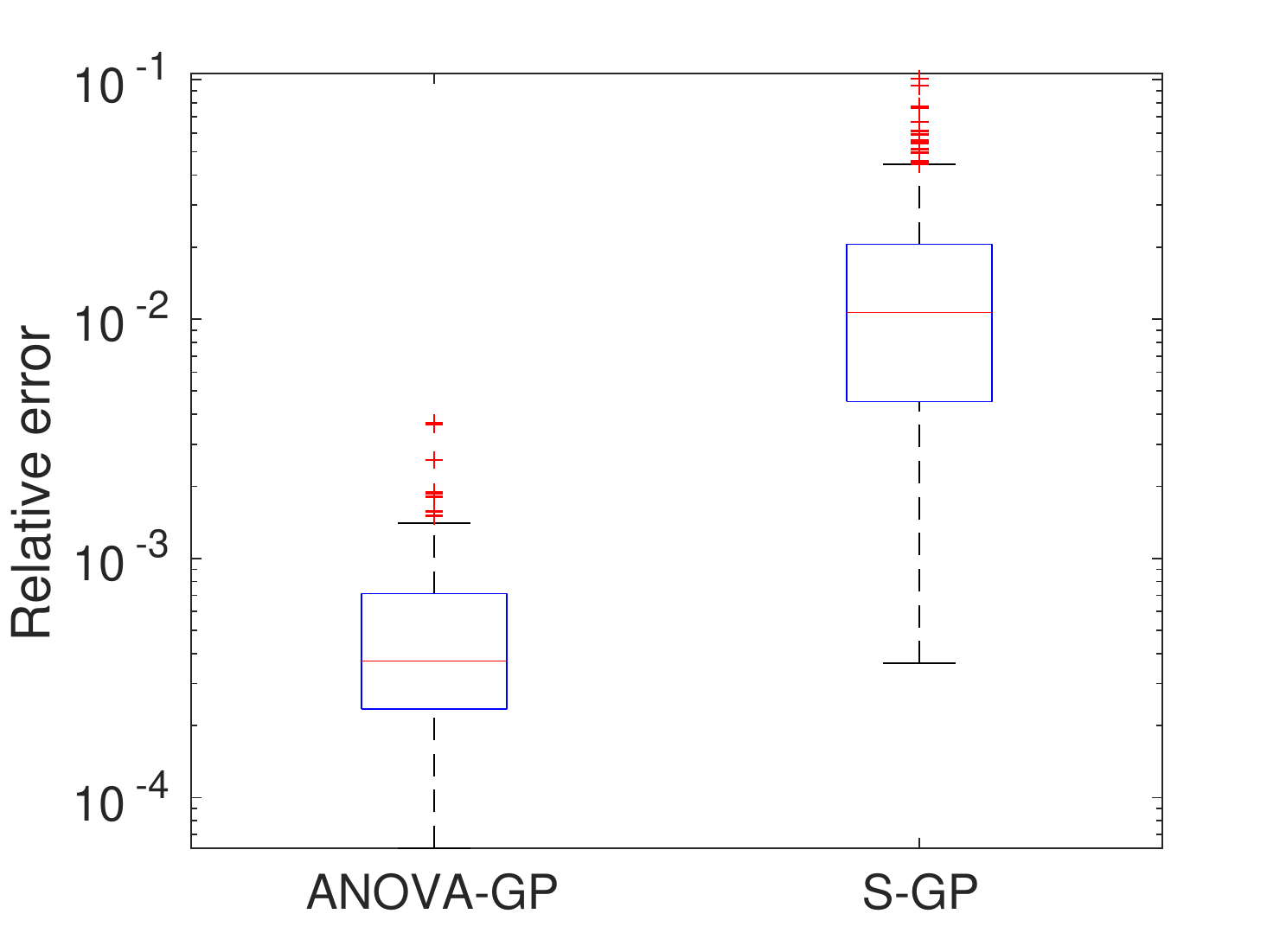}
	\caption{Relative errors of ANOVA-GP and S-GP for $200$ test data points, test problem 2.}
	\label{Box_sk_109}
\end{figure}

For this test problem, 
since the simulator output for the Stokes problem involves velocity and pressure approximations, 
the relative mean \eqref{weights}  
is defined to be the sum of the functional $L_2$ norms of the approximation  
functions associated with them, i.e.,
$
\gamma_t:={(\|\pEp(u_t)\|_{L_2}+\|\pEp(p_t)\|_{L_2})}/{(\|\sum_{s \in \pP_0\cup\cdots\cup \pP_{|t|-1}} \pEp(u_s)\|_{L_2}+\|\sum_{s \in \pP_0\cup\cdots\cup \pP_{|t|-1}} \pEp(p_s)\|_{L_2})},
$
where $u_t$ and $p_t$ denote ANOVA terms for velocity and pressure respectively (see \eqref{comfunc5}). 
The tolerance for selecting ANOVA terms is set to $tol_{index}=10^{-5}$, 
and the quadrature rule is set to the tensor products of 
one-dimensional Clenshaw-Curtis quadrature with five quadrature points in Algorithm \ref{AA},
while  the tolerance for PCA is set to $tol_{pca}=10^{-3}$ in Algorithm \ref{PCA}. 
In this setting, the index set $\pP$ constructed through Algorithm \ref{AA} only contains the zeroth order index and $32$
first order indices, i.e., $|\pP_1|=32$ and $|\pP|=33$. 
The number of training points for ANOVA-GP is set to $N_{train}=50$ 
(as the input of Algorithm \ref{AGP})
and that for S-GP is set to $N:=(|\pP|-1)\times N_{train}=1600$ (as the input of Algorithm \ref{SGP}) for a fair comparison. 
Again, 200 samples of $\xi$ are generated and the corresponding simulator outputs are computed. 
The errors of ANOVA-GP and S-GP are assessed through the relative error defined in \eqref{r_error}.
Figure \ref{Box_sk_109} shows errors for both ANOVA-GP and S-GP,
where it is clear that the errors ANOVA-GP are one order of magnitude smaller than the errors of S-GP. 
In addition, the number of principal components retained for S-GP is $30$ and that for each ANOVA term in ANOVA-GP is 
one, which indicates that the ANOVA terms \eqref{comfunc5} have very small ranks.

\begin{figure}[htpb!]
	\begin{multicols}{2}
		\centering  
		\subfigure[Streamline, simulator]{
			\label{fig_kl109_v_GT}
			\includegraphics[width=0.45\textwidth]{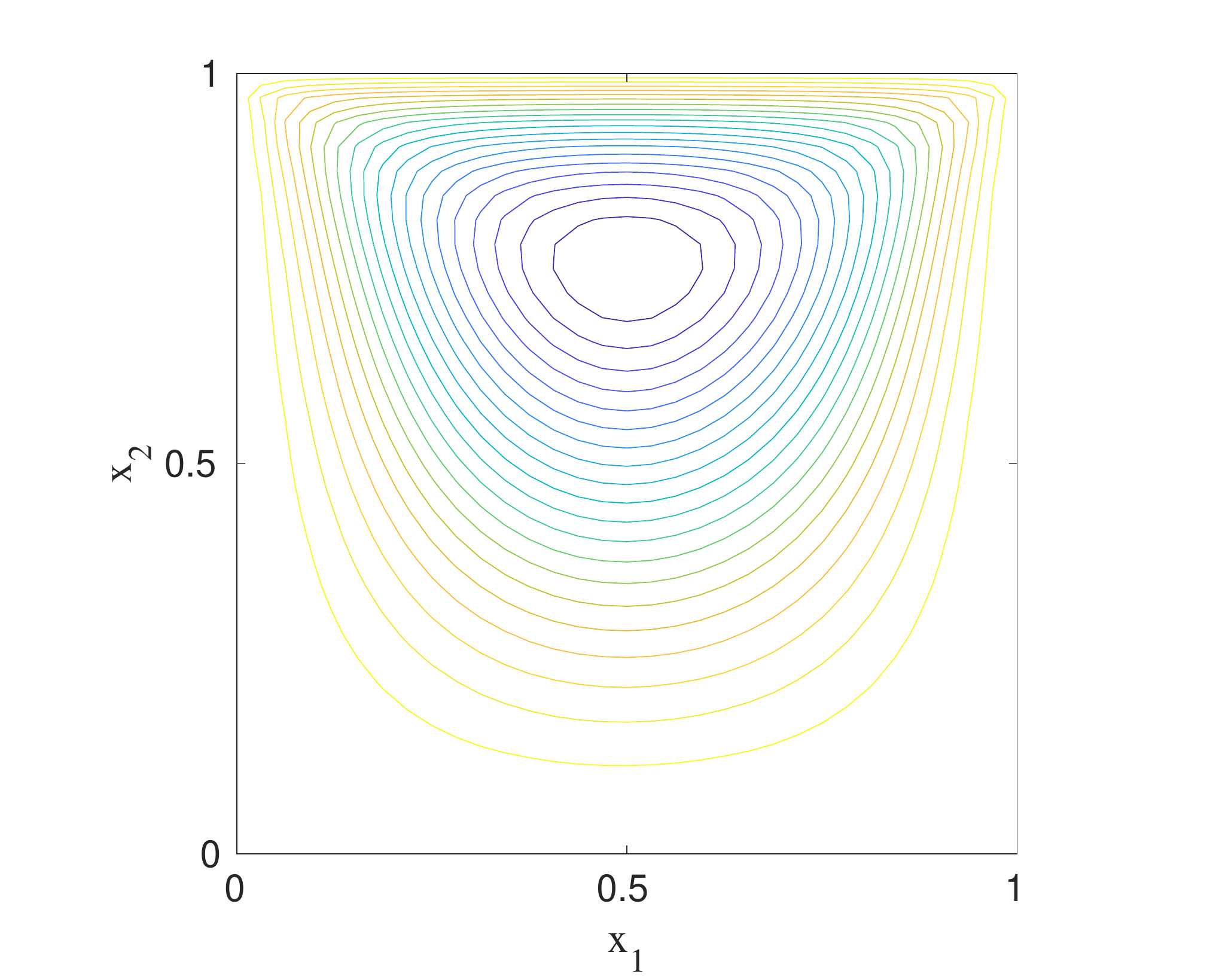}}
		\subfigure[Streamline, S-GP]{
			\label{fig_kl109_v_GPM}
			\includegraphics[width=0.45\textwidth]{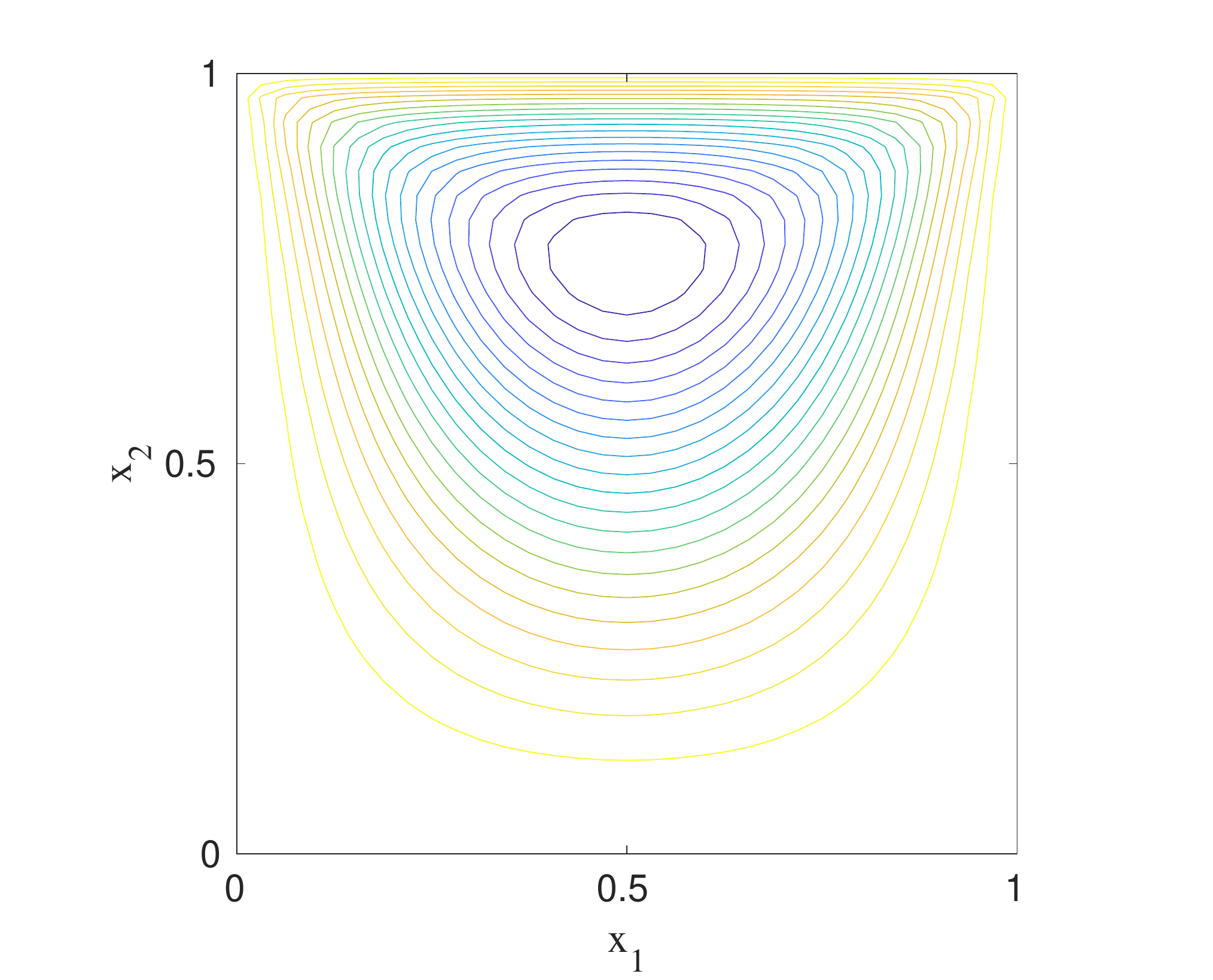}}
		\subfigure[Streamline, ANOVA-GP]{
			\label{fig_kl109_v_AT-GPM}
			\includegraphics[width=0.45\textwidth]{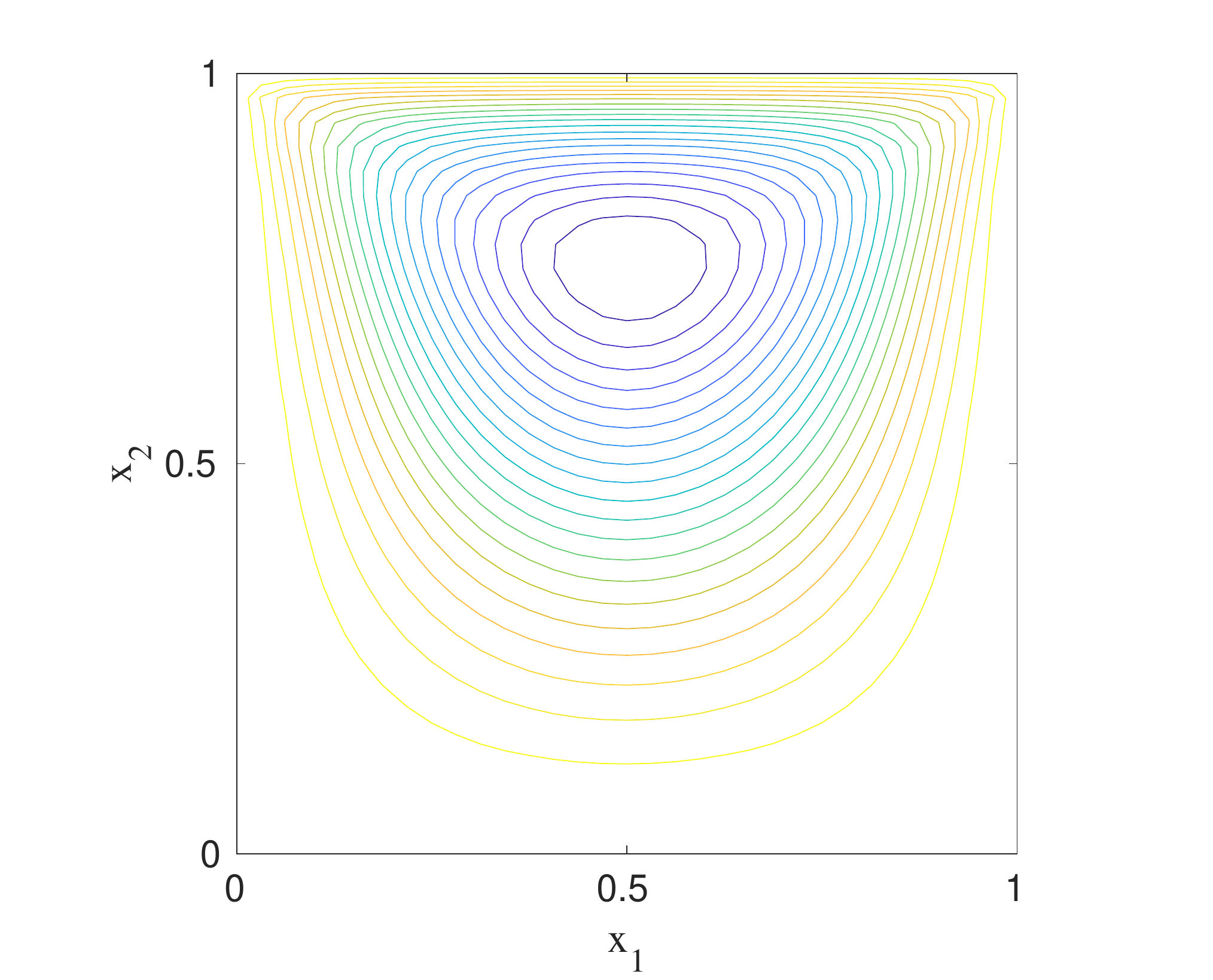}}
		\subfigure[Pressure, simulator]{
			\label{fig_kl109_p_GT}
			\includegraphics[width=0.45\textwidth]{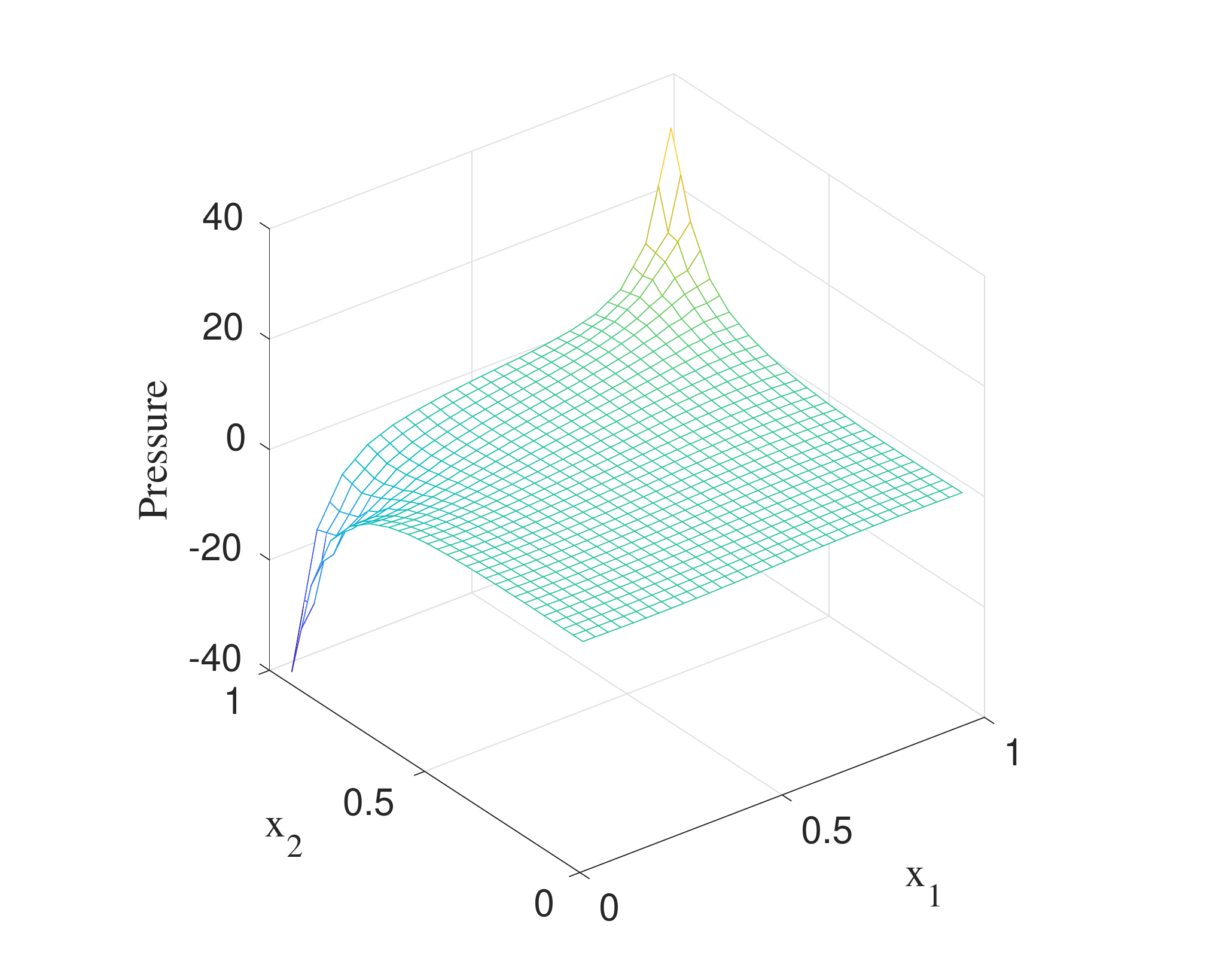}}
		\subfigure[Pressure, S-GP]{
			\label{fig_kl109_p_GPM}
			\includegraphics[width=0.45\textwidth]{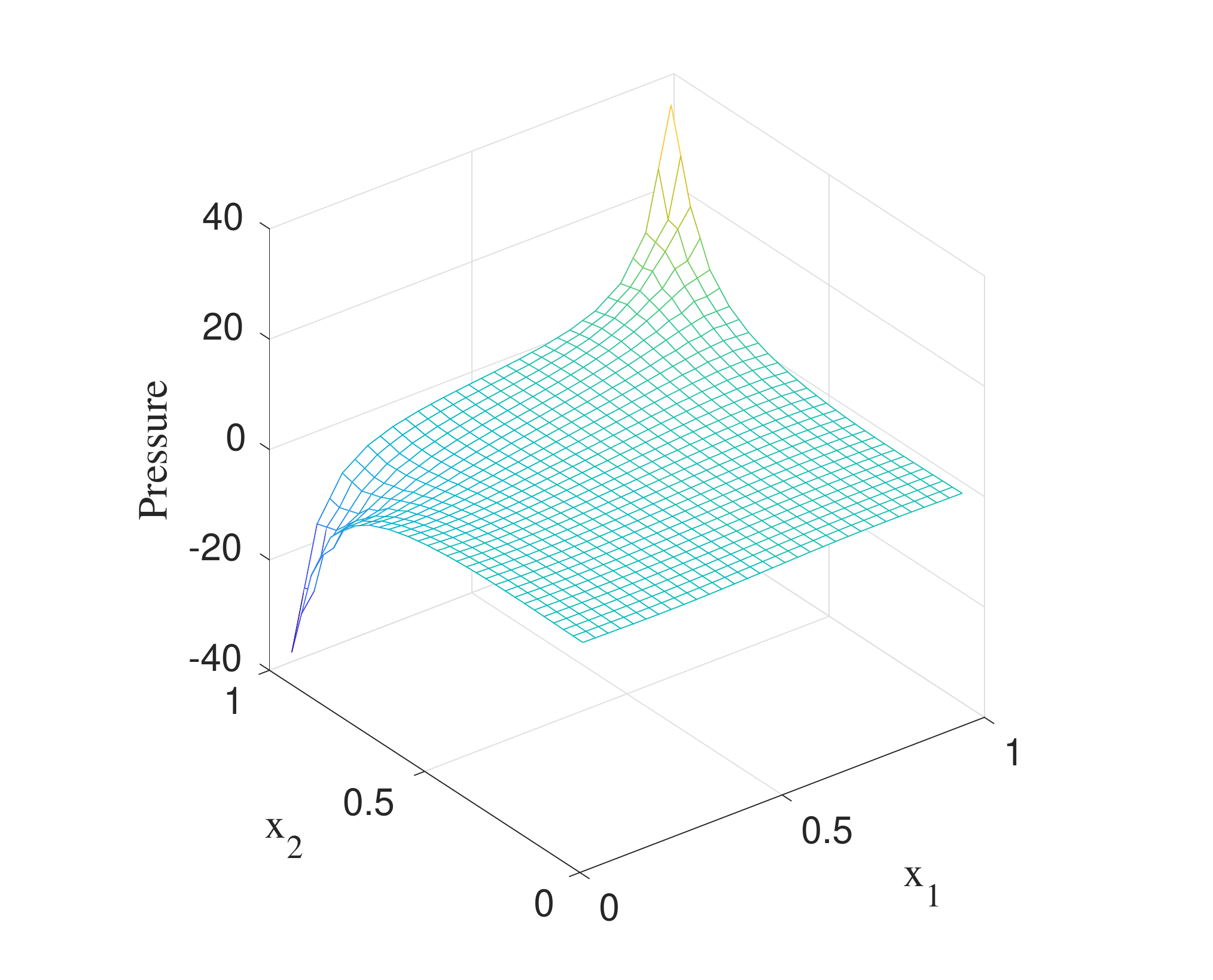}}
	          	\subfigure[Pressure, ANOVA-GP]{
			\label{fig_kl109_p_AT-GPM}
			\includegraphics[width=0.45\textwidth]{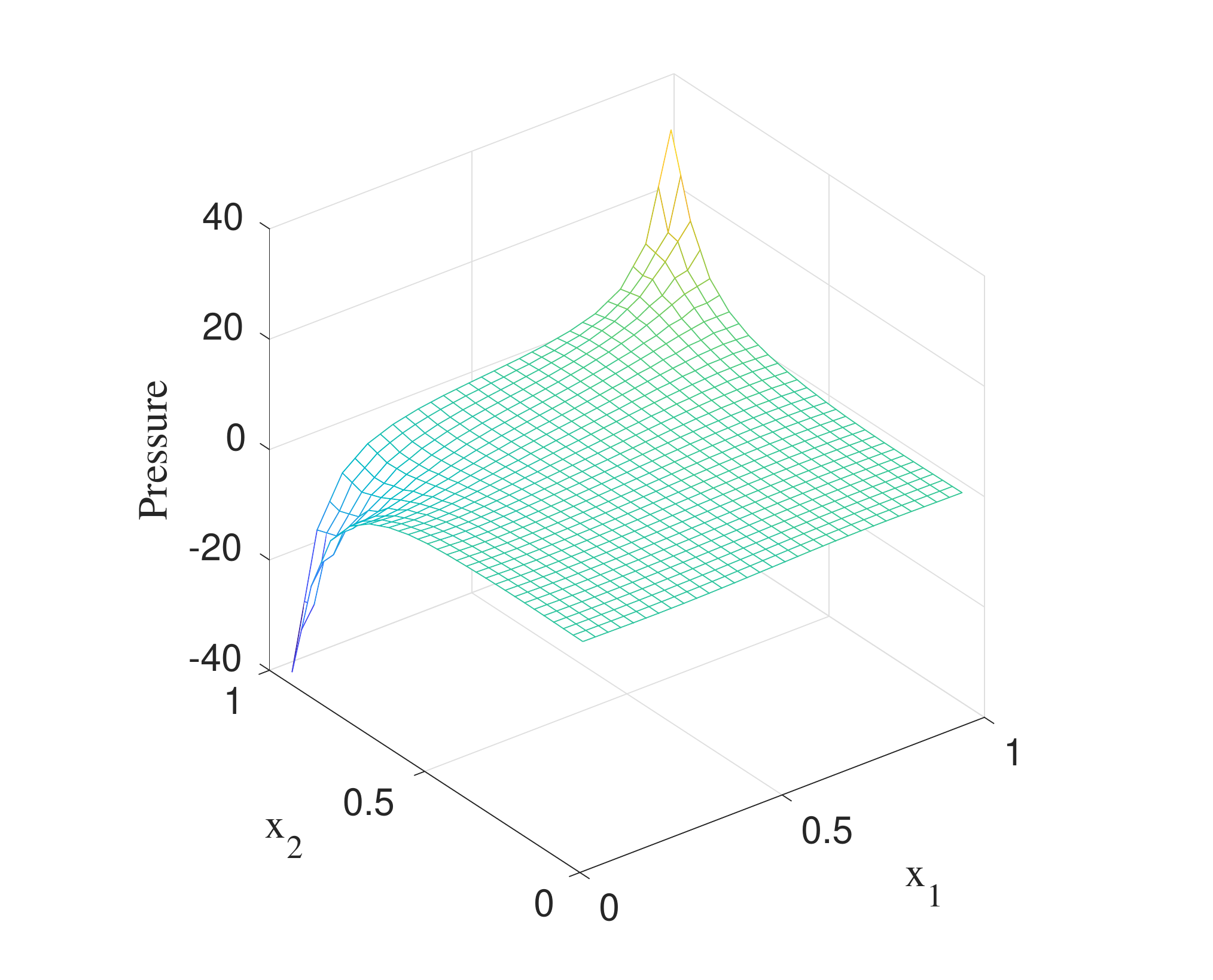}}
	\end{multicols}
	\caption{Examples of predictions made by S-GP and ANOVA-GP, and the simulator outputs, test problem 2.}
	\label{kl_109_example}
\end{figure}

\begin{figure}[htpb!]
	\begin{multicols}{2}
	\centering  
	\subfigure[Velocity error of S-GP]{
		\label{fig_kl109_v_GPM_EM}
		\includegraphics[width=0.45\textwidth]{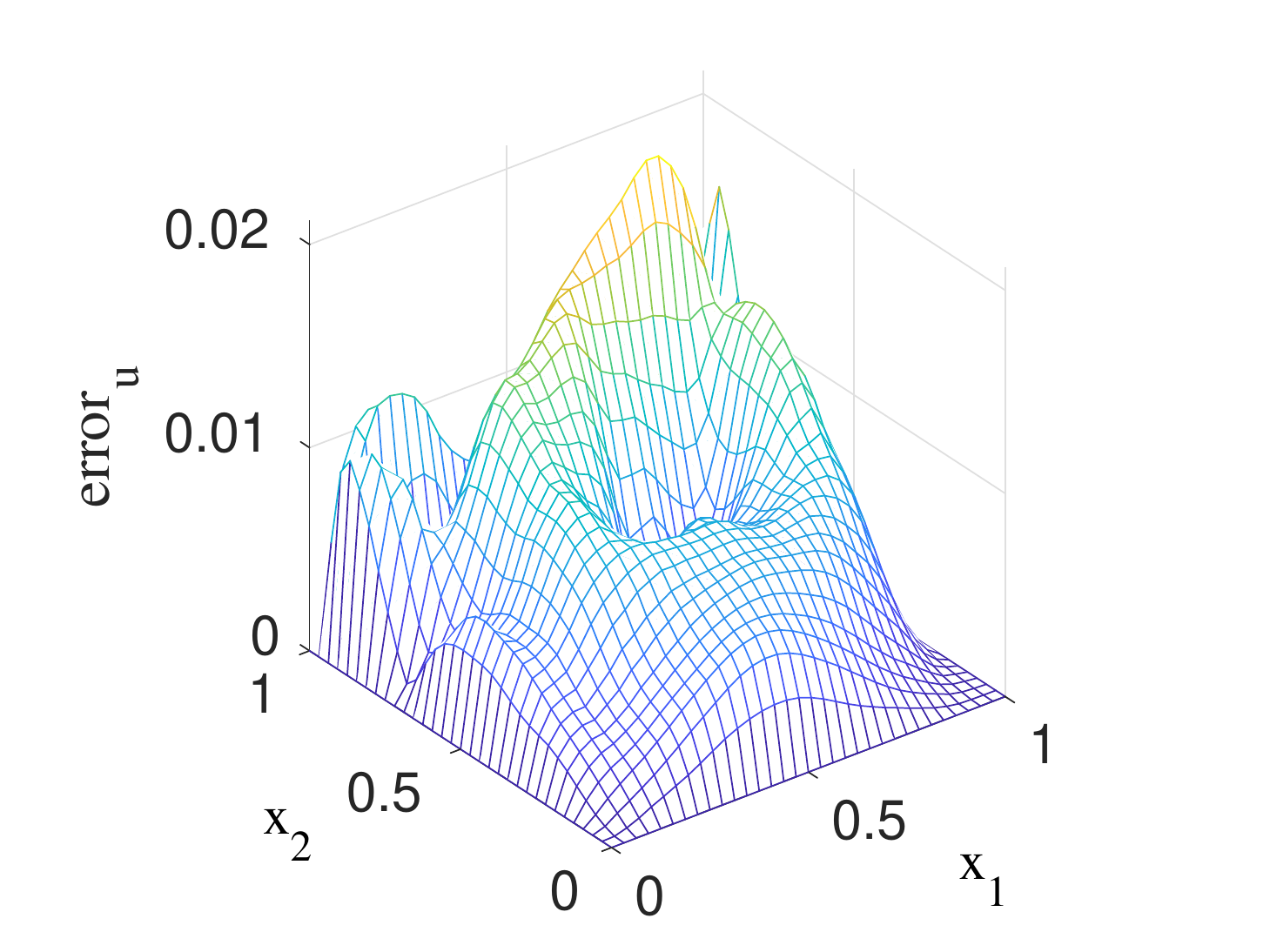}}
	\subfigure[Velocity error of ANOVA-GP]{
		\label{fig_kl109_v_AT-GPM_EM}
		\includegraphics[width=0.45\textwidth]{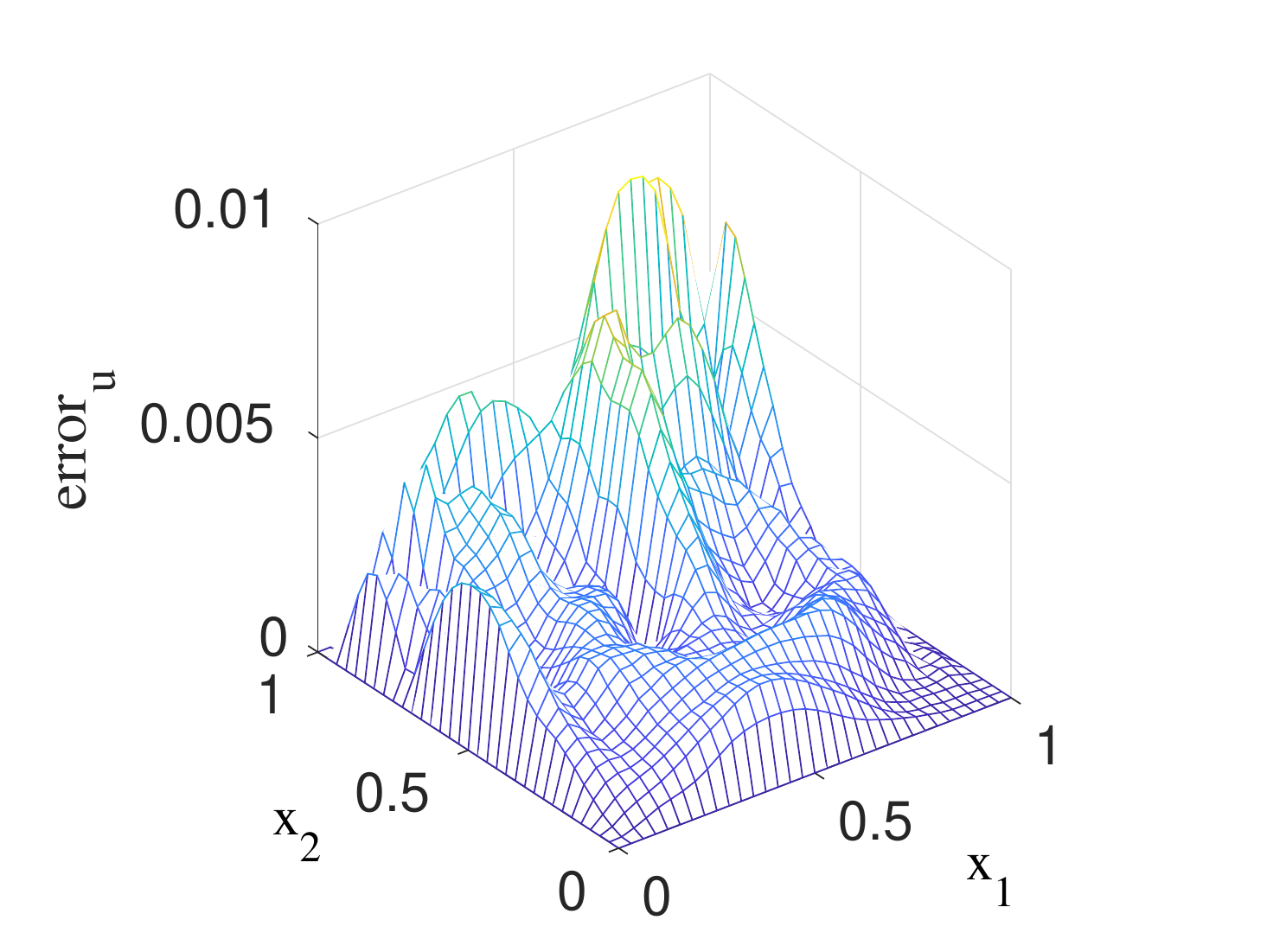}}
	\subfigure[Pressure error of S-GP]{
		\label{fig_kl109_p_GPM_EM}
		\includegraphics[width=0.45\textwidth]{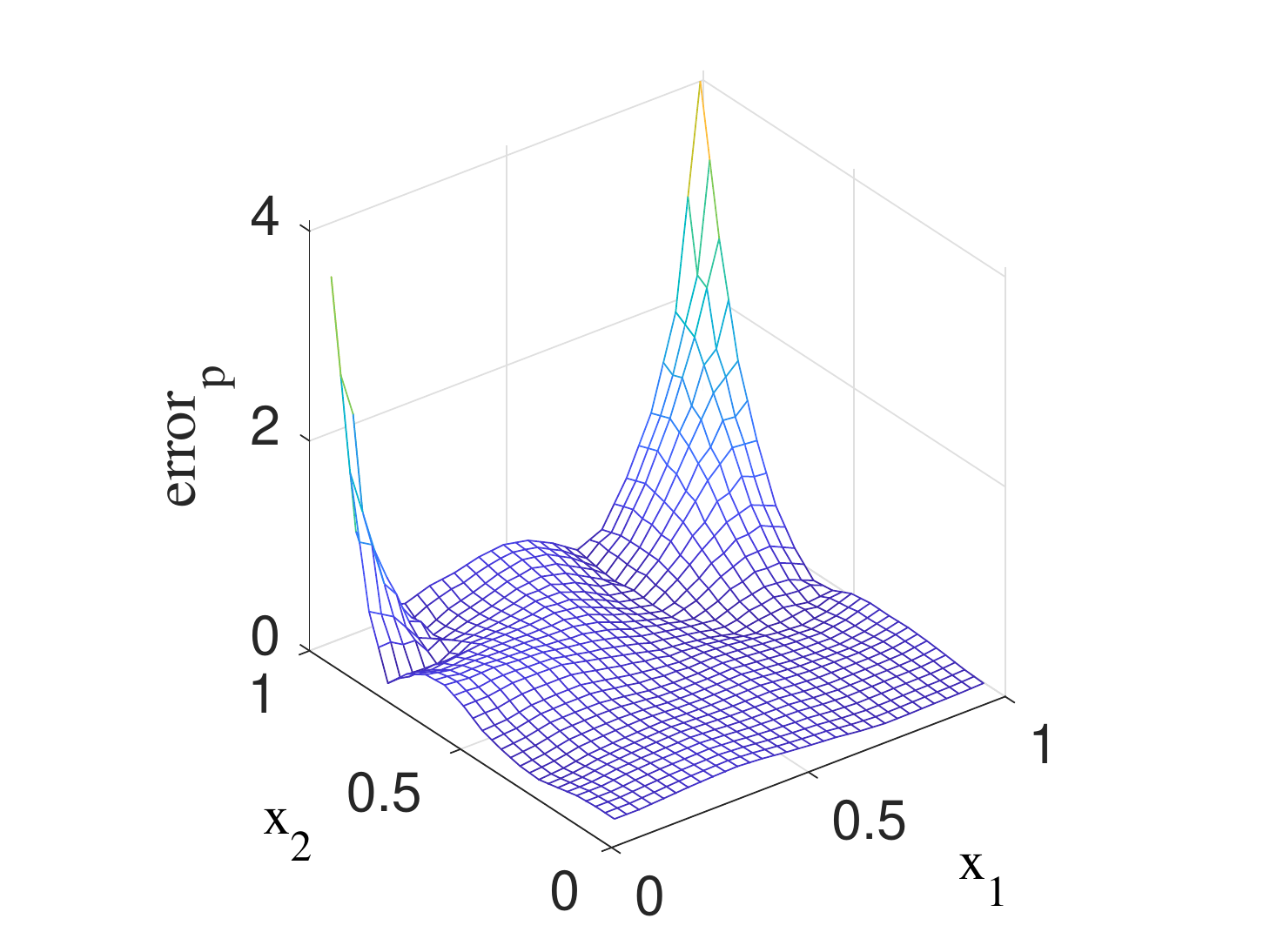}}
	\subfigure[Pressure error of ANOVA-GP]{
		\label{fig_kl109_p_AT-GPM_EM}
		\includegraphics[width=0.45\textwidth]{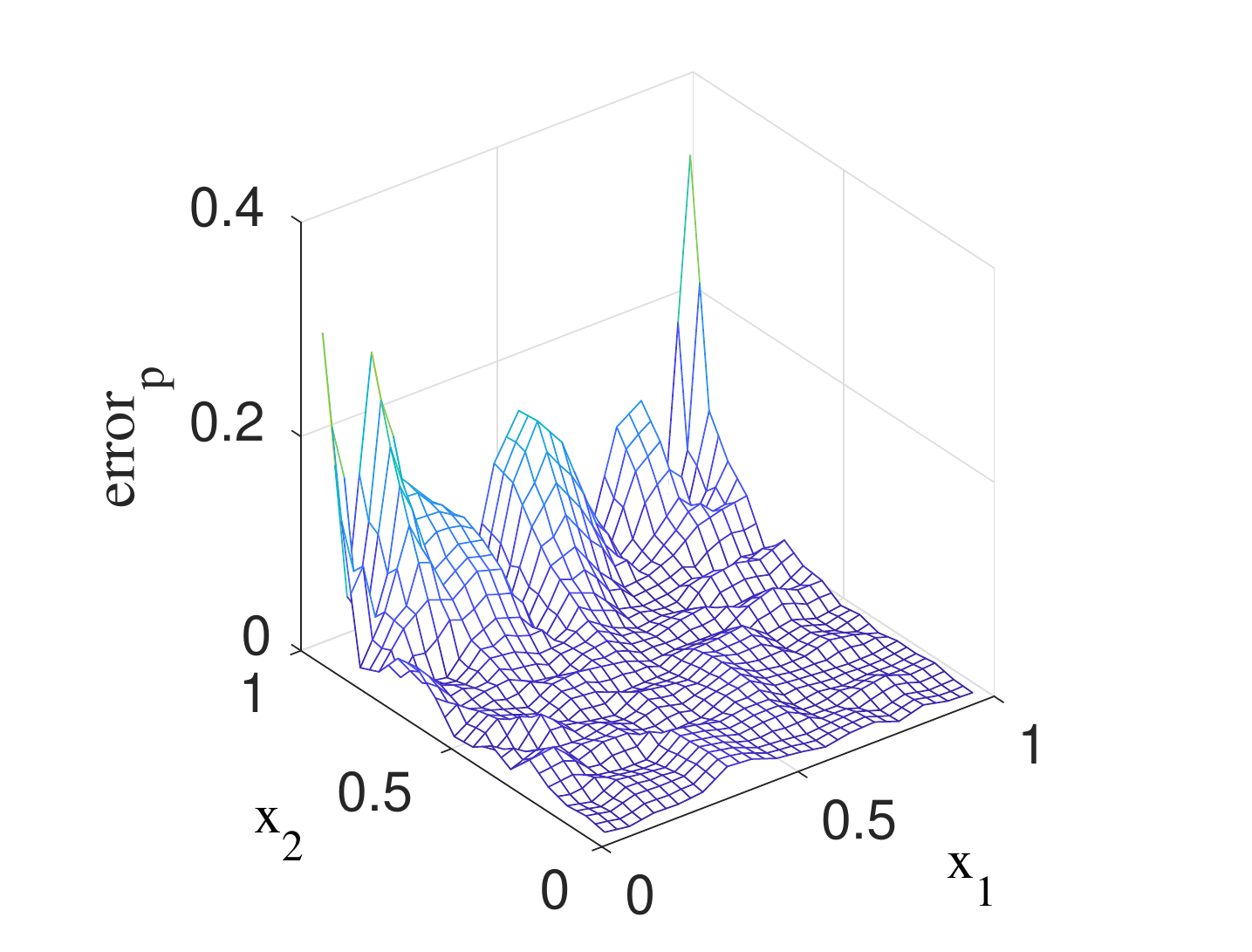}}
	\end{multicols}
	
	\caption{
	Errors of S-GP and ANOVA-GP predictions, test problem 2.
	}
	\label{kl_109_example_error_map}
\end{figure}

Figure \ref{kl_109_example} shows the simulator output and 
the ANOVA-GP and the S-GP predictive means  corresponding to a given
realization of $\xi$. 
From Figure \ref{kl_109_example}(a), Figure \ref{kl_109_example}(b) and Figure \ref{kl_109_example}(c),
it can be seen that the velocity streamlines
obtained from the simulator output and those from ANOVA-GP  and  S-GP emulators 
are visually indistinguishable. However, 
from \ref{kl_109_example}(d), Figure \ref{kl_109_example}(e) and Figure \ref{kl_109_example}(f),
the pressure obtained from S-GP is clearly larger than that of the simulator near the upper right corner (1,1),
while the pressure fields obtained from ANOVA-GP and the simulator are visually indistinguishable. 
To look more closely, we compute the errors of the  
emulator predictive means as follows.
For a physical grid point, let  $u=[u_1,u_2]^T$ and $p$ denote the velocity and the pressure 
obtained through the simulator at this grid point,
and $\overline{u}=[\overline{u}_1,\overline{u}_2]$ and $\overline{p}$ 
denote the velocity and the pressure obtained through the emulators (predictive means of 
ANOVA-GP and S-GP). 
The errors of velocity and pressure at this grid point are
defined as $error_u:= \sqrt{(u^2_1 -\overline{u}_1)^2+(u^2_2 -\overline{u}_2)^2}$
and $error_p = |p - \overline{p}|$ respectively. 
Figure \ref{kl_109_example_error_map} shows these errors.
From Figure \ref{kl_109_example_error_map}(a) and Figure \ref{kl_109_example_error_map}(b),
it can be seen that the maximum  velocity error of  ANOVA-GP is less than half of the maximum error 
of S-GP. From Figure \ref{kl_109_example_error_map}(c) and Figure \ref{kl_109_example_error_map}(d),
the maximum pressure error of ANOVA-GP is less than ten percent of the maximum error of S-GP.

